%% file: main.tex
\documentclass{aastex63}
\usepackage{amsmath}

\input{newcommandstex}

\received{January 26, 2023}
\revised{May 16, 2024}
\accepted{November 9, 2024}
\submitjournal{ApJ}
\shortauthors{Kashyap and Hanasoge}
\shorttitle{Mode coupling: Modelling Center-to-Limb systematic}

\begin{document}

\title{Modelling the Center-to-Limb systematic in normal-mode-coupling measurements}

\correspondingauthor{Samarth G. Kashyap}
\email{kashyap@mps.mpg.de}

\author[0000-0001-5443-5729]{Samarth G. Kashyap}
\affil{Tata Institute of Fundamental Research \\
Homi Bhabha Road, Navy Nagar, Mumbai - 400005, India}
\affil{Max-Planck-Institut f\"{u}r Sonnensystemforschung, \\
Justus-von-Liebig-Weg 3, 37077 G\"{o}ttingen, Germany}

\author[0000-0003-2896-1471]{Shravan M. Hanasoge}
\affil{Tata Institute of Fundamental Research \\
Homi Bhabha Road, Navy Nagar, Mumbai - 400005, India}
\affil{Center for Space Science, New York University Abu Dhabi, \\
PO Box 129188, Abu Dhabi, UAE}

\defcitealias{Woodard-2013-SoPh}{W13}
\defcitealias{Gizon-2020-Science}{G20}
\begin{abstract}
Solar meridional circulation, which manifests as poleward flow
near the surface, is a relatively weak flow. While
meridional circulation has been measured through various local
helioseismic techniques, there is a lack of consensus about the nature
of the depth profile and location of return flow,
owing to its small amplitude and poor signal-to-noise ratio in
observations.
The measurements are strongly hampered by systematic effects,
whose amplitudes are comparable to the signal induced by the flow
and modelling them is therefore crucial. The removal of the center-to-limb
systematic, which is the largest known feature hampering the inference of meridional
flow, has been heuristically performed in helioseismic analyses, but
it's effect on global modes is not fully understood or modelled.
Here, we propose both a way to model the center-to-limb systematic and
a method for estimation of meridional flow using global helioseismic
cross-spectral analysis.
We demonstrate that the systematic cannot be ignored while modelling the
mode-coupling cross-spectral measurement and thus is critical for the inference
of meridional circulation. We also show that inclusion of a model for the center-to-limb systematic improves shallow meridional circulation estimates from cross-spectral analysis.
\end{abstract}



\section{Introduction}
\label{sec:intro}
Meridional circulation (MC) is a relatively weak large-scale flow,
with a surface amplitude of $\sim$ 20 ms\({}^{-1}\)
\citep{Giles-1997-Natur, Basu-Antia-1999-ApJ,
Zhao-Kosovichev-2004-ApJ, Hanasoge-2022-LRSP}
compared to $\sim$ 200 - 300 ms\({}^{-1}\) for differential rotation
\citep[for e.g., ][]{Schou-1998-ApJ}.
MC is directed from the equator to the poles
in both the hemispheres and the polar accumulation of mass
is prevented by the inward plunging of the flow at high latitudes.
In the interior, the flow reverse direction
heading from the poles to the equator, with radial outflow back to the surface
at low latitudes.
The sustenance of differential rotation and the 11-year magnetic-activity cycle are among the long-standing puzzles in solar physics and MC is
one a critical ingredient in setting these dynamics
\citep{Dikpati-2009-odsm, Charbonneau-2010-LRSP,
Choudhuri-2021-SCPMA, Hanasoge-2022-LRSP}.
Thus, an understanding of the depth profile and latitudinal
structure of MC is of great interest to solar physics at large. \\

Surface MC has been observed for several decades using a multitude of techniques \citep[see][for a detailed review]{Hanasoge-2022-LRSP}, such as,
sunspot tracking \citep{Howard-Gilman-1986-ApJ},
tracking of photospheric magnetic features \citep[e.g.,][]{Komm-1993-SoPh} and
Doppler velocity measurements \citep[e.g.,][]{Hathaway-1993-ASPC}.
The study of the depth-profile of meridional flow has been possible through helioseismology.
Over the years, the most successful attempts
to estimate MC have predominantly used techniques of local helioseismology.
\cite{Giles-1997-Natur}
estimated sub-surface flow in the top 4\% of the Sun
using data from the \emph{Michelson Doppler Imager} (MDI) using time-distance helioseismology
and found MC to be constant throughout this range.
Subsequently, \cite{Giles-2000-PhDT} found MC to have a magnitude of 3~\(\mathrm{ms}^{-1}\)
at the base of the convection zone. Near-surface MC has also been
constrained to be 20~\(\mathrm{ms}^{-1}\) using ring-diagrams
\citep{Schou-1998-ApJL, Basu-Antia-1999-ApJ, Zhao-Kosovichev-2004-ApJ}.
\cite{Basu-Antia-1999-ApJ} used helioseismic data from MDI and found
no change in the sign of MC up to a depth of 21 Mm. \cite{Zhao-Kosovichev-2004-ApJ}
also used MDI data and measured a magnitude of 20 ms${}^{-1}$, up to a
depth of 12 Mm, using 6 years of data.
\cite{Zhao-2013-ApJL} found the MC to have a two-cell structure in
depth. \cite{Rajaguru-Antia-2015-ApJ} used 4 years of HMI observations,
imposing a mass-conservation constraint using stream functions and
found the return flow of MC to be near the base of the convection zone
at r=0.77$R_\odot$. \cite{Rajaguru-Antia-2020-ASSP} found that they
could obtain a single- or double-cell MC
depth profile depending on the frequencies of modes used for the seismic measurements.
The implications of double-cell MC on dynamo
models are discussed in \cite{Hazra-2014-ApJ}. More recently,
\cite{Gizon-2020-Science}(G20, hereafter) measured MC over two solar cycles, using
data from both HMI and MDI, and showed that time-averaged MC is
single-celled in each hemisphere, with a mid-latitude magnitude of
4~\(\mathrm{ms}^{-1}\).
The lack of consensus and
theoretical considerations motivates us to explore the measurement of MC using global modes.\\

A systematic center-to-limb variation (C2L) in
measurements of helioseismic travel times was reported first by \cite{Duvall-Hanasoge-2009-ASPC}.
\cite{Zhao-2012-ApJL} found
the systematic phase difference to be a function of heliocentric
radius \citep[a catalog of coordinate systems can be found in][]{Thompson-2006-AnA}.
Although the physical origin is not completely understood,
\cite{Baldner-Schou-2012-ApJL} hypothesized that it could be due to the asymmetry in convective outflows and inflows, resulting in a
phase-shift in the wavefield and they found qualitative
agreement with the measurements of \cite{Zhao-2012-ApJL}.
\cite{Zhao-2016-SoPh} concluded that foreshortening could not explain C2L,
and that the C2L may due to radiative transfer effects and/or upflow-downflow
asymmetry due to convection, with both effects resulting in phase shifts in the observed modes \citep{Zhao-2022-arXiv}.
The results of \cite{Baldner-Schou-2012-ApJL}, \cite{Zhao-2016-SoPh}
and \cite{Zhao-2022-arXiv} motivate our choice to model the C2L as
a phase factor that varies with heliocentric radius and mode frequency.
Whether this factor leaves a significant imprint on cross-spectral
measurements is an important question as there have been no clear answers from literature relating to the modelling or the effect of such a systematic bias.
While there have been numerous studies of MC using normal-mode coupling \citep{Woodard-2013-SoPh, Schad-2012-AN, Schad-2013-ApJL}, they do not account for C2L-bias.
\cite{Schad-2013-ApJL} also claim the effect of C2L-bias to be small for their cross-spectrum-based measurement for $\ell<100$.
Here, we quantify the impact of a C2L-bias on cross spectra and demonstrate that it is indeed the most important aspect to consider when imaging MC.\\

%
We describe the mode-coupling measurement technique in Section~\ref{sec:theory}.
Section~\ref{sec:c2l-model} focuses on the modelling of the C2L-bias in the
context of global modes. The measurable impact of MC on mode-coupling measurements is shown in Section~\ref{sec:meridional-circ}.
We describe the parameterization of C2L-bias in Section~\ref{sec:c2l-param}.
The recipe for setting up an inversion is discussed in Section~\ref{sec:cross-spectra}
through an analysis of how the C2L-bias and meridional circulation affect
cross-spectral measurements. Aspects of modelling and challenges in
inferring meridional circulation are discussed in Section~\ref{sec:discussions}.

\section{Theoretical Formulation}
\label{sec:theory}
Turbulent convection drives acoustic waves which form resonances (``normal modes") in the Sun.
The reference model of the Sun (model S) is spherically symmetric,
non-rotating, non-magnetic, adiabatic and static. The normal modes of the
reference model form a complete, orthonormal basis. Hence, any wavefield,
\(\bfxi(\bfr, t)\), can be expressed as a linear combination of these normal-mode eigenfunctions, i.e.,
\begin{equation}
    \bfxi(\bfr, t) = \sum_k \tilde a^k(t) \tilde\bfxi^k(\bfr)
    \exp(i\tilde\omega_k t),
\end{equation}
where $\tilde\omega_k$ is the characteristic frequency of mode $k$,
$\tilde\bfxi^k(\bfr)$ is the spatial eigenfunction, $\tilde a^k$
is the mode amplitude, where the tilde is used to refer to quantities corresponding
to the reference model.
The index $k$ is a compact notation to denote the index-tuple $(n, \ell, m)$, where $n$ denotes the radial order, $\ell$ denotes the spherical harmonic degree and $m$ is the azimuthal order. Whenever necessary, the indices are explicitly mentioned.
Using the Chandrasekhar-Kendall decomposition \citep{Chandrasekhar-Kendall-1957-ApJ}, we
write the eigenfunctions $\tilde\bfxi^{n\ell m}$ in terms of vector spherical harmonics $(\bfY^{\ell m}, \bfPsi^{\ell m}, \bfPhi^{\ell m})$ (see Appendix~\ref{apdx:vsh}) along with their vertical and horizontal
components $(U^{n\ell}, V^{n\ell})$ respectively,
\begin{equation}
    \tilde\bfxi^{n\ell m}(r, \theta, \phi) =
    U^{n\ell} \bfY^{\ell m}(\theta, \phi) +
    V^{n\ell}(r) \bfPsi^{\ell m}(\theta, \phi),
\end{equation}
where $\theta$ denotes co-latitude and $\phi$ denotes longitude. Since we observe only the line-of-sight
component of the wavefield on about half the solar surface, a perfect
decomposition of the wavefield into spherical harmonics is not possible. The line-of-sight projection and windowing in space are quantified using
``leakage matrices" $\tilde{L}^{\ell m}_{\ell' m'}$ \citep[for e.g., ][]{Schou-Brown-1994-A&AS, Howe-Thompson-1998-A&AS}  as
\begin{equation}
    \varphi^{n \ell m}(\omega) =
    \tilde L^{n \ell m}_{n \ell' m'}
    \tilde  a^{n \ell' m'}(\omega),
\end{equation}
where $\varphi^{n\ell m}(\omega)$ is the Fourier-counterpart of
the observed spherical-harmonic time series.
Perturbations (such as flows) to the reference model result in
changes to the characteristic frequencies $\omega_k$ and distortion of eigenfunctions, given by
\begin{equation}
    \tilde\omega_k \to \omega_k \qquad
    \tilde\bfxi^k(\bfr) \to \bfxi^k(\bfr) =  \sum_{k'} c^k_{k'}\tilde\bfxi^{k'}(\bfr),
\end{equation}
where $\omega_k$ is the characteristic frequency of the mode
in the presence of a perturbation,
$\bfxi^k$ represents the distorted eigenfunctions and $c^k_{k'}$
denote the coupling-coefficients.

\subsection{Cross-spectral model}
The primary observable in mode coupling is the cross-spectrum, which is expressed generally as the
expectation value of the cross-spectrum $\langle \varphi^{n \ell m}(\omega) \varphi^{*n' \ell' m'}(\omega') \rangle$, where $\langle ... \rangle$ is used to denote the statistical expectation.
Temporally stationary perturbations only affect cross-spectra
across the same frequency channel, i.e., $\omega = \omega'$ and axisymmetric perturbations dominantly influence $m=m'$ cross-spectra \citep{Lavely-Ritzwoller-1992-RSPTA}. In general mode-coupling measurements have sufficient signal when the modes in question are closely located in frequency-space.
A complete analysis would also need to include considerations of cross-radial couplings, i.e., $n \neq n'$. However, we restrict the present analysis to couplings of the same radial order $n=n'$ for two reasons: (a) cross-couplings of different radial orders are relatively small in number, owing to their distance in frequency-space, and (b) the theory of cross-$n$ mode-coupling is not fully developed yet since they may possibly more susceptible to systematical errors and influence of background power. Therefore, for brevity, we omit the radial order $n$ in all subsequent expressions.
\begin{equation}
  \label{eqn:cross-spectra-model}
  \langle \varphi^{\ell m} \varphi^{*\ell' m'} \rangle = \sum_{\ell_1 m_1 \ell_2 m_2 \ell_3 m_3}
  \tilde{L}^{\ell m}_{\ell_1 m_1} \tilde{L}^{\ell' m'}_{\ell_2 m_2} c^{\ell_1 m_1}_{\ell_3 m_3}
  c^{*\ell_2 m_2}_{\ell_3 m_3} |a^{\ell_3 m_3}|^2.
\end{equation}
The computed cross-spectra, when plotted as a function of $m$ and $\omega$
appear as a slanted set of parallel lines. A stack-summing operation
\citep[for details see, ][]{SGK-2021-ApJS} is performed to improve the
signal-to-noise ratio (SNR). For illustration, the stacked cross-spectrum
is seen in the left-panel of Figure~\ref{fig:p4-cs-sample-nonzerot} and
the corresponding stack-summed spectra are shown on the right-panel.

\subsection{Coupling coefficients due to meridional circulation}
\label{sec:meridional-circ}
We implement the formalism
proposed by \cite{Vorontsov-2011-MNRAS} and use it in the same manner as in \cite{SGK-2021-ApJS}. The flow is written in terms of vector spherical harmonics. For an axisymmetric flow such as meridional circulation, we obtain 
\begin{equation}
    \bfu(\bfr) = \sum_{s} 
    \left[ 
    u_{s}(r) \bfY^{s0}(\theta, \phi) + 
    v_{s}(r) \bfPsi^{s0}(\theta, \phi) + 
    w_{s}(r) \bfPhi^{s0}(\theta, \phi)
    \right],
    \label{eqn:flow-perturbation-sph-expansions}
\end{equation}
where $u_s(r), v_s(r)$ are poloidal flow components and $w_s(r)$ is the toroidal component. Rotation is toroidal (i.e., no radial flow component) and meridional circulation is a poloidal flow (both lateral and radial flows.)
The coupling coefficients $c^i_j$ are
written in terms of the distance between the angular degrees of the two modes $p=\ell - \ell'$. The expressions are reproduced from \cite{Vorontsov-2011-MNRAS} for clarity's sake.
\begin{equation}
    c_{\ell}^{\ell+p} = \tfrac{1}{\pi} \int_0^{\pi} \cos \left[pt - \sum_{k=1,2,...}\tfrac{2}{k}\mathrm{Re}(b_k^{n\ell}) \sin{(kt)} \right] \times \mathrm{exp}\left[i \sum_{k=1,2,...} \tfrac{2}{k} \mathrm{Im}(b_k^{n\ell}) \cos{(kt)} \right] \mathrm{d}t, \qquad p=0,\pm1,...
    \label{eqn:clp},
\end{equation}
where $b_k$ are written in terms of generalized $a-$coefficients, $a_s^k$. 
The real and imaginary parts of $b_k$ are written in terms of $a_s^k$ as follows
\begin{eqnarray}
    \mathrm{Re}(b_k^{n\ell}) &=& \ell \left(\frac{\partial \tilde{\omega}}{\partial \ell} \right)^{-1}_n
    \sum_{s+k = \mathrm{odd}}
    P_s^k \left(\frac{m}{\ell} \right)
    \mathrm{Re}[a_s^k]^{n\ell},
    \quad k = 1,2,... \label{eqn:b_k_real} \\
    \mathrm{Im}(b_k^{n\ell}) &=& \ell
    \left(\frac{\partial \tilde{\omega}}{\partial \ell} \right)^{-1}_n
    \sum_{s+k = \mathrm{even}} 
    P_s^k\left(\frac{m}{\ell} \right) 
    \mathrm{Im}[a_s^k]^{n\ell}, \quad k = 1,2,... \label{eqn:b_k_imag}.
\end{eqnarray}
Coefficients $a_s^k$ correspond to kernel-averaged flow velocities, given by
\begin{align}
    \mathrm{Re}[a_s^k]^{n\ell} &= (-1)^{\frac{s-k+1}{2}}
    \frac{(s-k)!!(s+k)!!}{(s+k)!}
    \left(\frac{2s+1}{4\pi}\right)^{1/2} 
    \left\langle \frac{w_s(r)}{r} \right\rangle^{n\ell}, 
    \label{eqn:ask-real}\\
    \mathrm{Im}[a_s^k]^{n\ell} &= (-1)^{\frac{s-k+2}{2}}
     k\frac{(s-k-1)!!(s+k-1)!!}{(s+k)!} 
     \left(\frac{2s+1}{4\pi}\right)^{1/2}
     \left\langle \frac{v_s(r)}{r} \right\rangle^{n\ell}.
    \label{eqn:ask-imag}
\end{align}
Here, the averaging symbol \(\langle .. \rangle\) denotes kernel-averaged
quantities
\begin{equation}
    \langle \Gamma_s(r) \rangle^{n\ell} = \int_{\odot} dr \, \, \Gamma_s(r) \, \,
    \rho_0(r) r^2 \left( U^{n\ell}(r)^2 + \ell(\ell+1)V^{n\ell}(r)^2\right).
    \label{eqn:depth-averaged}
\end{equation}
Using the kernels for averaging the velocity profiles given by Eqn.~(\ref{eqn:depth-averaged}) over depth, we compute the real and imaginary parts of $b_k^{n\ell}$ using Eqns.~(\ref{eqn:b_k_real}, \ref{eqn:b_k_imag}). These components are used to compute the coupling-coefficients $c_\ell^{\ell+p}$ using Eqn.~(\ref{eqn:clp}), which are subsequently used for computing the cross-spectrum from Eqn.~(\ref{eqn:cross-spectra-model}).
\section{Modelling the C2L-bias}
\label{sec:c2l-model}
Following the modelling procedure of \cite{Baldner-Schou-2012-ApJL},
we introduce an additional phase factor to the observed velocity field on the surface. The spherical harmonic transform (SHT) of the observed wavefield on the surface
$\varphi(\theta, \phi, \tau)$ is denoted by $\varphi^{\ell m}$
\begin{equation}
    \varphi^{\ell m}(\tau) = \int_\odot d\Omega \,
    \varphi(\theta, \phi, \tau) \, Y^{\ell m *}(\theta, \phi)
    \label{eq:wavefield-sht},
\end{equation}
where $\tau$ is time. The C2L-bias may also be modelled as a shift in the phase of the ``true" wavefield, which is quantified using a phase factor $\exp(i f(r_\rmH))$, where $r_\rmH$ is the heliocentric radial distance from the disk-center of the Sun
\citep[defined as $\rho$ in Section 3.2, ][]{Thompson-2006-AnA}.
Let us denote the true and observed wavefields as
$\varPhi(\theta, \phi, \tau)$ and $\varphi(\theta, \phi, \tau)$, respectively. We then have
\begin{equation}
    \varphi^{\ell m}(\tau) = \int_\odot d\Omega\,
    \varphi(\theta, \phi, \tau) \, Y^{\ell m*} (\theta, \phi)=
    \int_\odot d\Omega \,
    \varPhi(\theta, \phi, \tau) \exp(if(r_\mathrm{H})) \, Y^{\ell m *}(\theta, \phi)
    \label{eq:obs-wavefield-sht}.
\end{equation}
Expanding both the true wavefield and the C2L-bias in terms of spherical
harmonics,
\begin{equation}
    \varPhi(\theta, \phi, \tau) = \sum_{\ell' m'} \varPhi^{\ell' m'}(\tau)
    Y^{\ell' m'}, \qquad \exp(if(r_\rmH)) = \sum_{st} \Lambda^{st} Y^{st}
    \label{eq:true-wavefield-sht}.
\end{equation}
It may be shown that (see Appendix~\ref{sec:deriv-leakage}) the above relations may be used to derive a new leakage matrix that encodes the C2L-bias as well,
\begin{equation}
    L^{\ell m}_{\ell_1 m_1} = (-1)^{m} \gamma_\ell \sum_{\ell' s}
       \gamma_s \gamma_{\ell'}
     \wig{\ell'}{s}{\ell}{0}{0}{0}
     \sum_{m' t}\Lambda^{st}
     \wig{\ell'}{s}{\ell}{m'}{t}{-m}
     \tilde{L}^{\ell' m'}_{\ell_1 m_1},
     \label{eq:modified-leakage}
\end{equation}
where $\tilde{L}^{\ell' m'}_{\ell_1 m_1}$ is the standard leakage matrix
that captures the effects of
line-of-sight projection, spatial windowing and apodization, and
$L^{\ell m}_{\ell_1 m_1}$ denotes the leakage matrix that
also includes the C2L-bias model, $\Lambda_{st}$. The standard leakage matrices
$\tilde{L}^{\ell m}_{\ell_1 m_1}$ are purely real whereas the modified leakage
matrices $L^{\ell' m'}_{\ell_1 m_1}$ are complex
because $\Lambda_{st}$ introduces an imaginary component (i.e., phase).
Using the new leakage matrices, the cross-spectral model can now be written as
\begin{equation}
  \label{eqn:p4-mod-cross-spectra-model}
  \langle \varphi^{\ell m}(\omega) \varphi^{*\ell' m'}(\omega) \rangle = \sum_{\ell_1 m_1 \ell_2 m_2 \ell_3 m_3}
  L^{\ell m}_{\ell_1 m_1} L^{\ell' m'}_{\ell_2 m_2} c^{\ell_1 m_1}_{\ell_3 m_3}
  c^{*\ell_2 m_2}_{\ell_3 m_3} |a^{\ell_3 m_3}(\omega)|^2.
\end{equation}
The stack-summed model-spectra, $\mcM$, are defined as
\begin{equation}
  \label{eqn:p4-mod-stack-summed}
  \mcM^{\ell \ell'}_{nt}(\omega) = \sum_m \mcS_m\langle \varphi^{\ell m}(\omega) \varphi^{*\ell' m+t}(\omega) \rangle,
\end{equation}
where $\mcS_m$ denotes the stack-summing process, where peaks of spectra at different azimuthal order $m$ are stacked and summed, in order to improve signal-to-noise ratio (for details, see \cite{SGK-2021-ApJS}).
\begin{figure}
    \centering
    \includegraphics[width=0.9\textwidth]{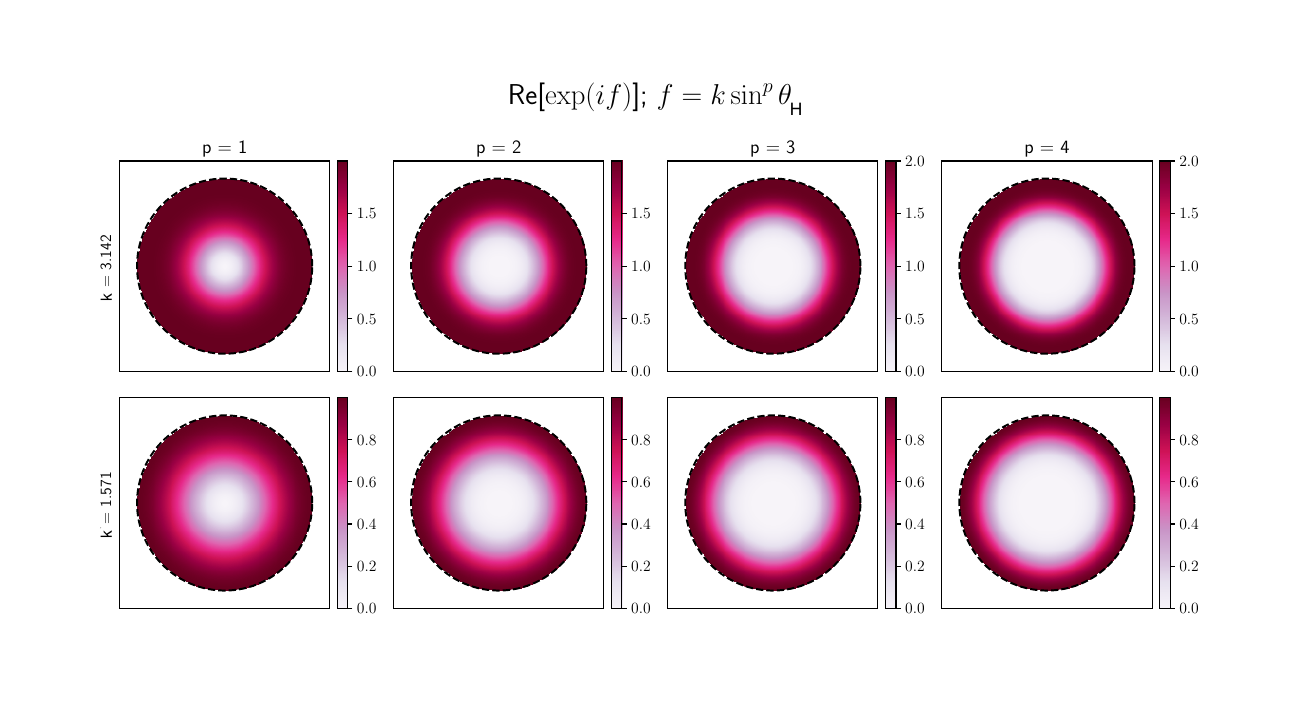}
    \caption{Plot of different C2L-bias profiles based on $\sin\theta_H$
    parameterization.}
    \label{fig:p4-true-ctol}
\end{figure}
\begin{figure}
    \centering
    \includegraphics[width=0.9\textwidth]{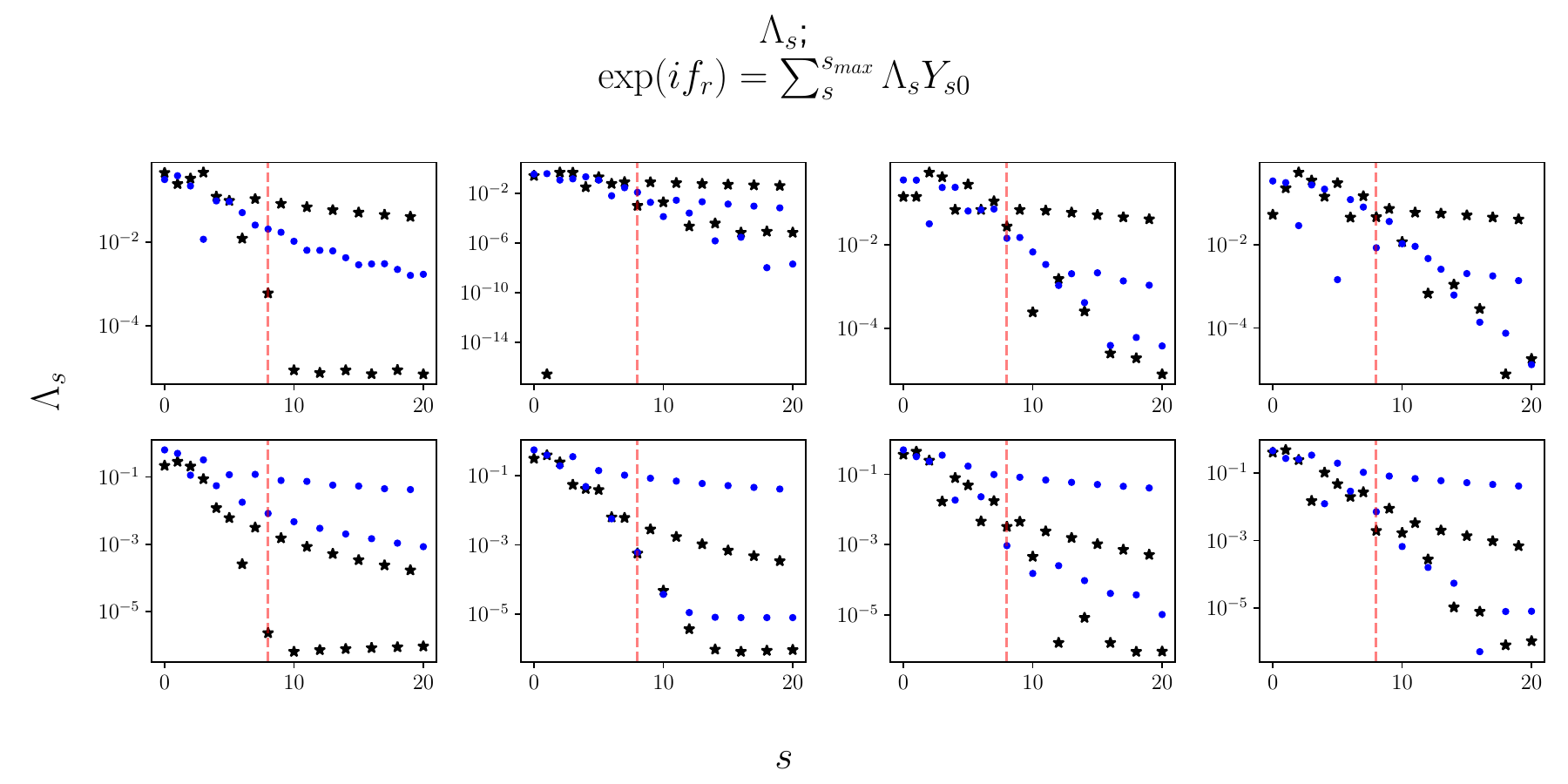}
    \caption{The C2L bias, given by $\exp(if(r_\rmH))$ is expanded in terms of spherical harmonics as
    $\sum_s^{s_\mathrm{max}} \Lambda_s Y_{s0}$. The stars indicate real part of $\Lambda_s$ and
    dots indicate the imaginary part. The y-axis is in logarithmic scale. The dashed-red line shows the value of $s$ at which the spherical harmonic expansion is truncated.}
    \label{fig:p4-lambda-s}
\end{figure}
\begin{figure}
    \centering
    \includegraphics[width=0.9\textwidth]{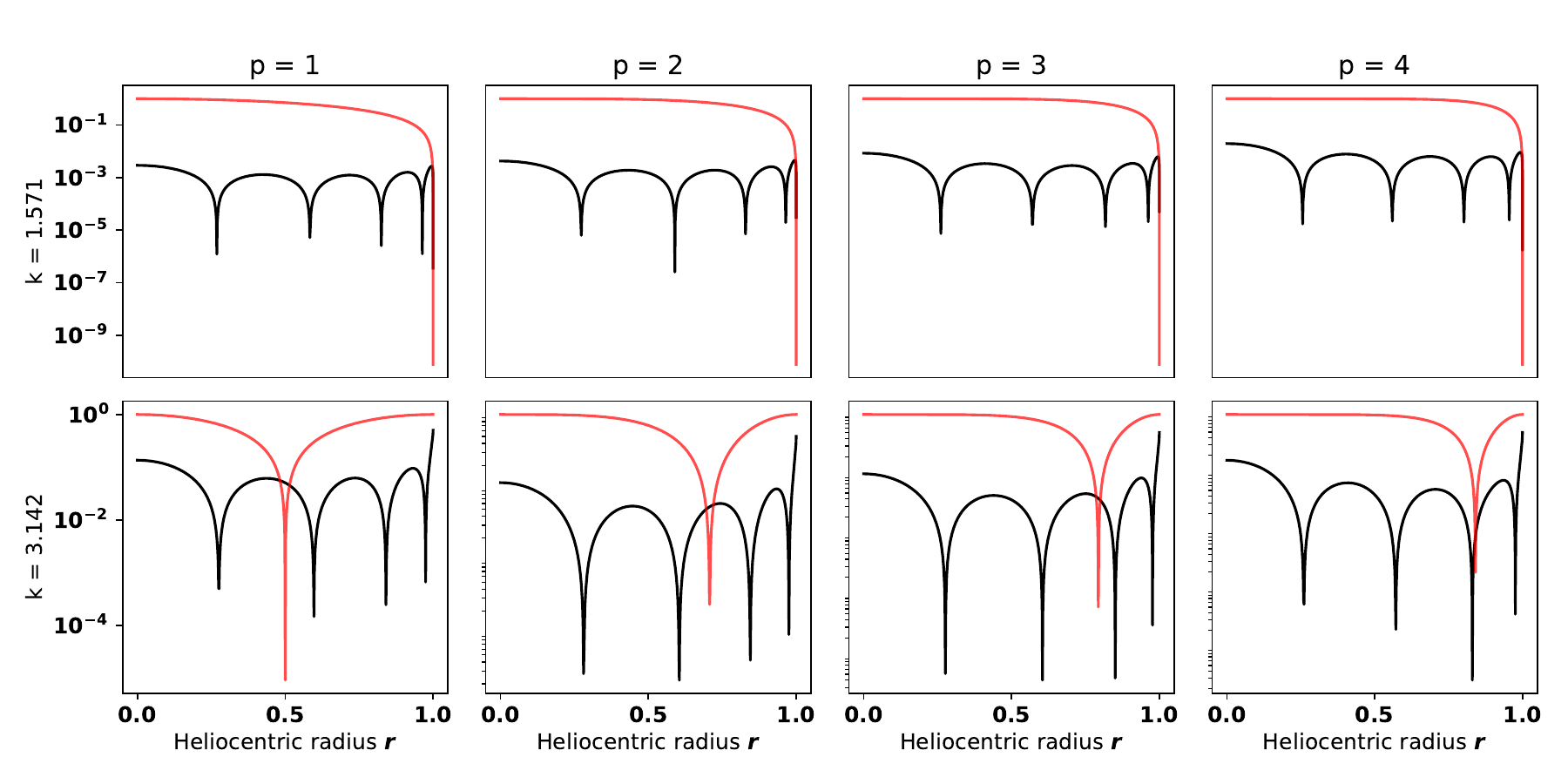}
    \caption{Error in estimating the C2L bias due to the truncation of the
    spherical-harmonic expansion at $s_\mathrm{max} = 7$, plotted in logarithmic scale. The true C2L-bias profile is shown as a solid-red line. The black line shows the error due to truncation of the series-expansion. In all the locations where the C2L-bias profile is non-negligible, the error due to truncation of spherical harmonic expansion is insignificant.}
    \label{fig:p4-diff-lineplot}
\end{figure}

\section{Paramteriztion of C2L-bias}
\label{sec:c2l-param}
For simplicity, we first
assume that the C2L phase factor is only a slowly varying
function of $r_\rmH$, which we express as a
polynomial in $r_\rmH$. For slowly varying phase, only the
first few powers of the polynomial in $r_\rmH$ are relevant. For greater precision, a more general basis function such as the B-spline would be preferred.
Here however, we choose a simple functional
form motivated by the following reasons:
(1) it is a good representation for slowly varying functions, and,
(2) given that the cross-spectral SNR is quite low, we prefer
to start with a cruder set of functions in order to establish
the method and study the consequences before using a more general expression. The C2L profile may be expressed as
\begin{equation}
    \Lambda(r_\rmH) = \exp \left( i \sum_{j} g_j r_\rmH^j\right).
    \label{eqn:c2l-gj-rj}
\end{equation}
This may also be written in the spherical-harmonic basis as
\begin{equation}
  \Lambda(r_\rmH) = \Lambda(\theta, \phi) = \sum_{st}\Lambda_{st}Y^{st}(\theta, \phi).
\end{equation}
The coefficients $\Lambda_{st}$,
where $(s, t)$ correspond to spherical-harmonic degree and
azimuthal order respectively, are used to describe the phase factor.
For a given spherical harmonic degree $s$, there
are $2s+1$ coefficients associated coefficients. Hence, if we consider parameters upto
$s=s_\mathrm{max}$, then we would have a total number $(s_\mathrm{max}+1)^2$
parameters. This translates to 36 parameters even for $s_\mathrm{max}=5$.
The number of parameters may be reduced drastically by using an
appropriate choice of basis. Since the C2L phase variation is purely a function of
heliocentric radius, in the reference where the disk-center of
the Sun corresponds to the pole of the coordinate system,
the C2L is an axisymmetric feature. In this coordinate system,
$\Lambda_{st} = 0$ for $t \neq 0$, due to axisymmetry. This implies that we need $s_\mathrm{max}+1$ parameters to represent the C2L systematic (details are described in
Appendix~\ref{sec:apdx:parameterization}). We also have that $\Lambda_{00} = 0$, which states that the constant phase shift across the disk is zero and hence does not change the cross
spectral measurement, leading to further reduction in the
number of representative parameters. Because the C2L bias is a phase, it imposes constraints on the
spherical-harmonic coefficients that are used to represent the
systematic, given by
\begin{equation}
    \int d\Omega \, | \Lambda(r)|^2 = 4\pi = \sum_s |\Lambda_s|^2.
    \label{eqn:lambda-constraint}
\end{equation}
This constraint acts as a normalization factor for the coefficients that represent the C2L-bias.
Figure~\ref{fig:p4-true-ctol}
shows C2L profiles for various degrees of polynomial $r_\rmH = \sin\theta_H$. The spherical-harmonic coefficients $\Lambda_{s0}$ are shown in Figure~\ref{fig:p4-lambda-s}, and as desired, the coefficients fall off rapidly as a function of harmonic degree $s$, indicating the C2L bias is a large-scale feature and is constructed
using low-order polynomials of $r_\rmH$. We reconstruct it using only the
$\Lambda_s$ values to the left of the truncation line (dashed-red) line in
Figure~\ref{fig:p4-lambda-s}. The difference between the reconstructed
and ``true" C2L profiles is shown in Figure~\ref{fig:p4-diff-lineplot}, indicating that
truncation is justified as it results in a high-fidelity reproduction.
\begin{figure}
    \centering
    \includegraphics[width=0.9\textwidth]{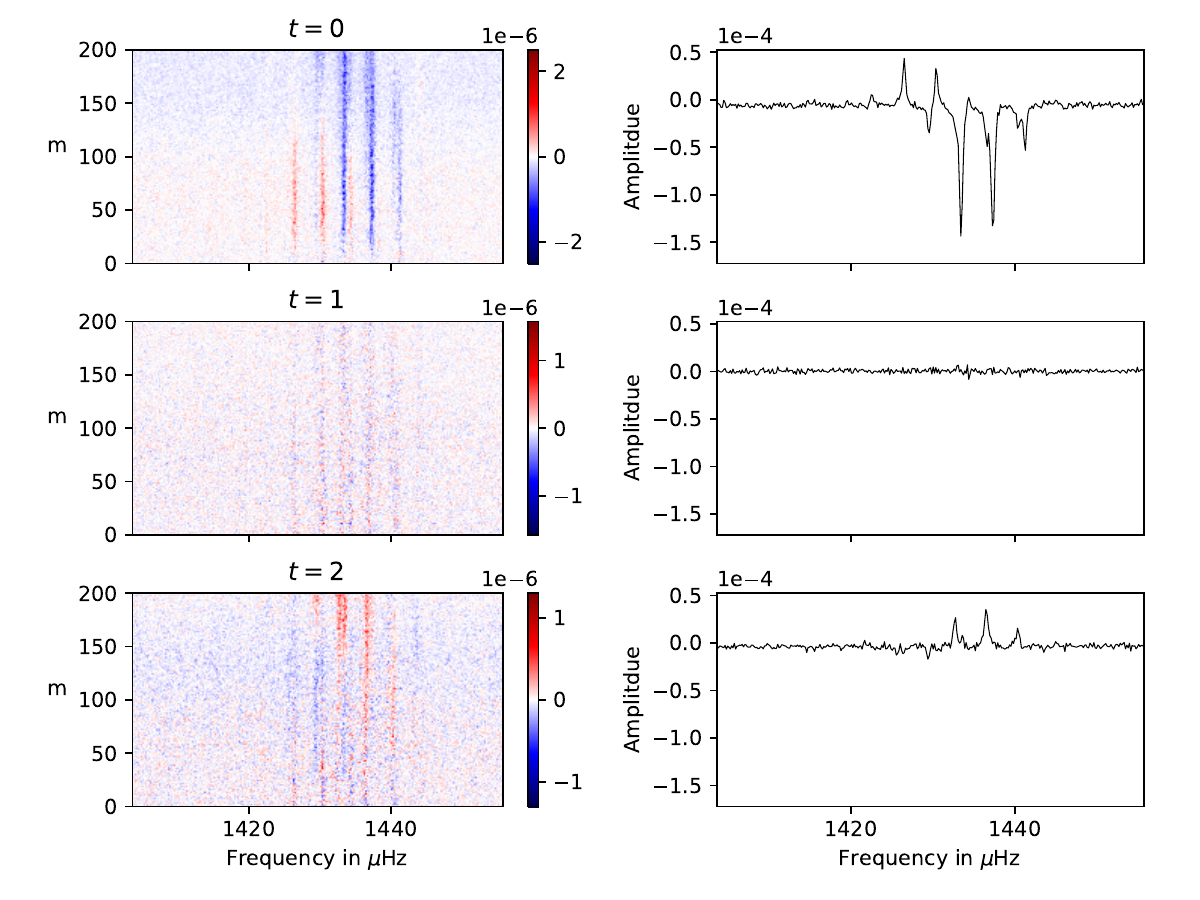}
    \caption{Real part of the cross-spectrum of modes $(0, 200)$ and $(0, 202)$ at finite $t$. The
    panels on the left show the stacked cross-spectral measurement
    $S^m(\omega)  =  \langle \varphi^{200, m}(\omega) \varphi^{202, m+t}(\omega) \rangle $
    for $m\ge 0$. The panels on the right correspond to stack-summed cross-spectra $\sum_m S^m(\omega)$.}
    \label{fig:p4-cs-sample-nonzerot}
\end{figure}
\section{Effect of C2L on cross-spectral observations and recipe for inversions}
\label{sec:cross-spectra}
MC solely affects the imaginary part of the cross-spectrum $\langle \varphi^{\ell m} \varphi^{* \ell' m} \rangle$ \citep[][W13, hereafter]{Woodard-2013-SoPh}.
Additionally, since MC is axisymmetric, it results in the coupling of modes where $m=m'$. However, we have seen in the previous
section that C2L may be represented using $\Lambda_{st}$, which may have
non-zero values for different $(s, t)$ pairs. This implies that the C2L signal
also modifies spectra at finite $t=m-m'$, through the modified leakage-matrix $L^{\ell m}_{\ell' m'}$. So far, attempts at estimating MC have involved the fitting of only $t=0$ spectra. The argument presented above makes a case for using $t \neq 0$  in addition to the $t=0$ spectra and provides us with extra data for simultaneously fitting the MC and C2L-bias parameters. Finite-$t$ cross-spectra are also affected by axisymmetric features through leakage. We compute finite-$t$
spectra in order to understand their utility in the estimation of MC through their constraints on the C2L profile. Based on the inspection of spectra across $(n, \ell)$ values, we present the following observations (see Figure~\ref{fig:p4-cs-sample-nonzerot} as well):

\begin{itemize}
  \item The cross-spectral signal at $t=0$ is the strongest. --- This channel contains contributions from the power-spectrum through leakage as well as from differential rotation and meridional circulation, the latter being axisymmetric $t=0$ features.
  \item Spectra in the even-$t$ channels are
  statistically significant although smaller in magnitude when compared with the $t=0$ channel. The $t\neq 0$ channels are expected to be smaller in magnitude as they contain cintributions from non-axisymmetric features and axisymmetric features through leakage and the C2L-bias.
  \item Spectra in odd-$t$ channels are smaller in magnitude than their even-$t$ counterparts. This is also a $t\neq 0$ channel but the lack of cross-spectral signal is surprising as there are no selection rules that indicate the lack of seismic signal in this channel. It is demonstrated later that the small, yet statistically significant features could be attributed predominantly to mode-coupling due to meridional circulation, i.e., the odd-$t$ channel needs to be included in MC estimation.
\end{itemize}
\subsection{Understanding odd-$t$ spectra}
\begin{figure}
    \centering
    \includegraphics[width=0.9\textwidth]{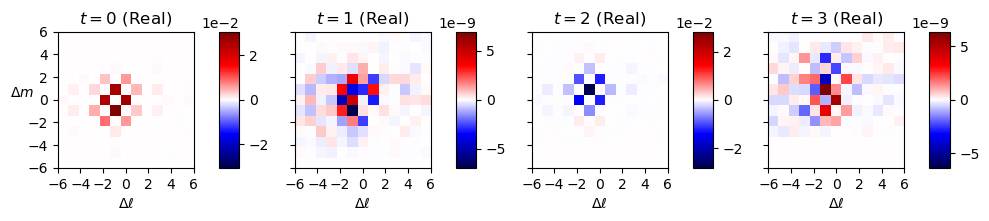}
    \caption{Product of leakage matrices in the expression of the cross-spectral model (Eqn.~\ref{eqn:cross-spectra-model}). To understand the observation in Figure~\ref{fig:p4-cs-sample-nonzerot} where the cross-spectral signal in $t=1$ channel is tiny, we show a slice of the product of leakage matrices
    $\tilde{L}^{\ell m}_{\ell_1 m_1} \tilde{L}^{\ell' m'}_{\ell_2 m_2}$. For illustration, we show the product for $\ell, \ell' = (200, 202)$, for different values of azimuthal difference $t=m-m'$, plotted for $m=20$. $\ell_1 = \ell, m_1=m$ and the vertical-axis represents $\Delta m = m_2 - m_1$ and the horizontal-axis represents $\Delta\ell=\ell_2 - \ell_1$. The product is small for odd-$t$.}
    \label{fig:p4-leak-prod}
\end{figure}

\begin{figure}
    \centering
    \includegraphics[width=0.9\textwidth]{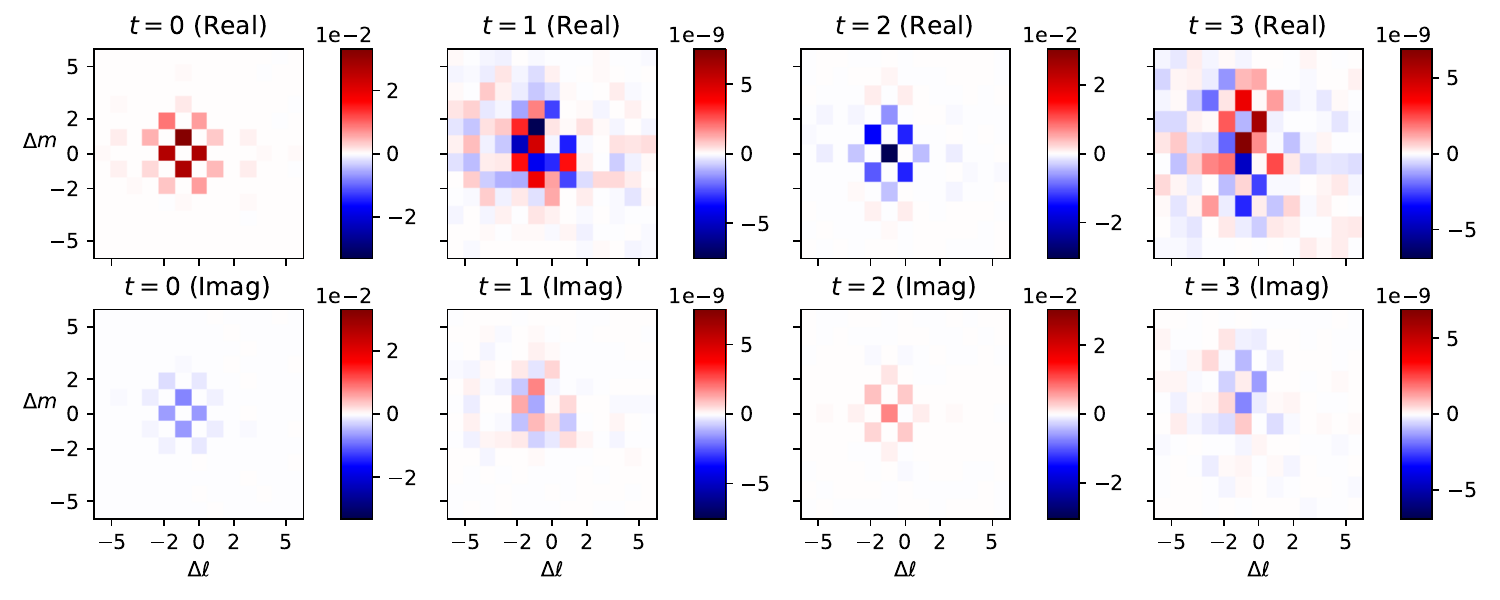}
    \caption{Similar to Figure~\ref{fig:p4-leak-prod}, but the leakage matrices here include the C2L-bias. The modified leakage matrices now have both real and imaginary components. Top panels contain real components while the bottom row shows the imaginary components.}
    \label{fig:p4-leakmod-prod}
\end{figure}

Since there are no selection rules to indicate a preference for the significance of odd- or even-$t$ spectra, we investigate terms in the cross-spectral model, Eqn.~(\ref{eqn:cross-spectra-model}). The factors can be categorized as those pertaining to leakage ($\tilde{L}^{\ell m}_{\ell_1 m_1} \tilde{L}^{\ell' m'}_{\ell_2 m_2}$), coupling ($c^{\ell_1 m_1}_{\ell_3 m_3} c^{*\ell_2 m_2}_{\ell_3 m_3} $)
and mode-profile ($|a^{\ell_3 m_3}|^2$). The mode-profile is a Lorentzian, irrespective of the measurement channel and hence cannot be attributed to the vanishing signal. The leakage term is a product of two leakage matrices. We computed the product of leakage matrices for odd- and even-$t$, a sample calculation of which is shown in Figure~\ref{fig:p4-leak-prod}. Indeed, we find that leakage-matrix products for odd $t$ are negligible by 7 orders of magnitude than those of the even-$t$ counterparts.
Hence, we conclude that even-$t$ cross-spectra are useful in the estimation of MC. However, it can also be seen in the middle-panel of Figure~\ref{fig:p4-cs-sample-nonzerot} that the odd-$t$ spectrum is small compared to even-$t$ but is not purely noise, as faint vertical streaks can be visually seen. To understand whether this is due to coupling or C2L, we study the leakage term using the C2L-modified leakage matrices $L^{\ell m}_{\ell_1 m_1} L^{\ell' m'}_{\ell_2 m_2}$. A sample of this is shown in Figure~\ref{fig:p4-leakmod-prod}, which shows the same trend as previously seen, i.e., the odd-$t$ leakage terms are 7 orders of magnitude smaller than their even-$t$ counterparts. Hence, we conclude that the odd-$t$ spectra are not purely noise due to the mode-coupling term, thus establishing the need to also use this data-channel for MC estimation.

\subsection{Forward models of cross-spectra}
\begin{figure}
    \centering
    \includegraphics[width=0.5\textwidth]{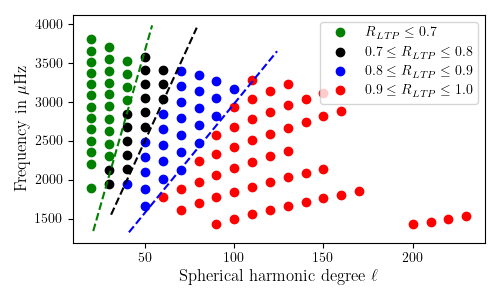}
    \caption{Set of modes used for forward model calculations. Modes are selected to span a range of frequencies and spherical harmonic degrees. Modes with $\nu/\ell<30 \mu$Hz are highlighted in red -- all of them have lower turning points $r_{LTP}$ between $0.9R_\odot$ and the surface. Modes shown in blue have $0.8 \le r_{LTP}/R_\odot \le 0.9$ and the modes shown in black have $0.7 \le r_{LTP}/R_\odot \le 0.8$. The green, black and blue lines are used to demarcate different regions. These lines are reproduced in other figures for ease of interpretability.}
    \label{fig:nu-by-ell}
\end{figure}

\begin{figure}
  \centering
  \includegraphics[width=0.5\textwidth]{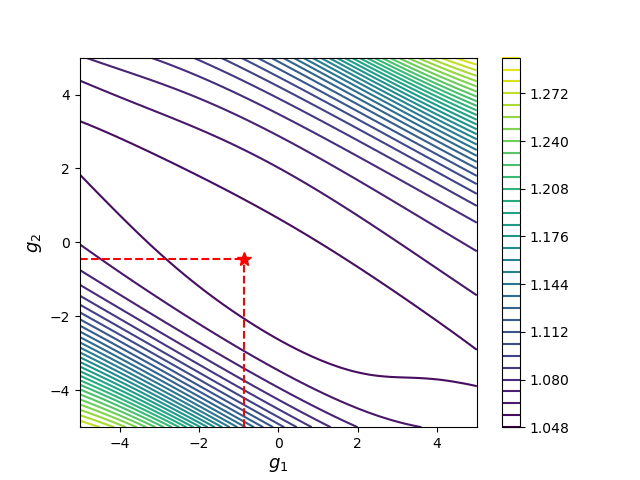}
  \caption{The reduced-$\chi^2$ misfit computed for different C2L-bias profiles modelled using 2 coefficients $(g_1, g_2)$. For this case, meridional circulation is set to 0 and the imaginary part of the cross-spectrum is completely attributable to the imaginary part of leakage matrix from the C2L-systematic. The red-star marks the minimum of the misfit function.}
  \label{fig:chisq-grid}
\end{figure}

\begin{figure}
  \centering
  \includegraphics[width=0.7\textwidth]{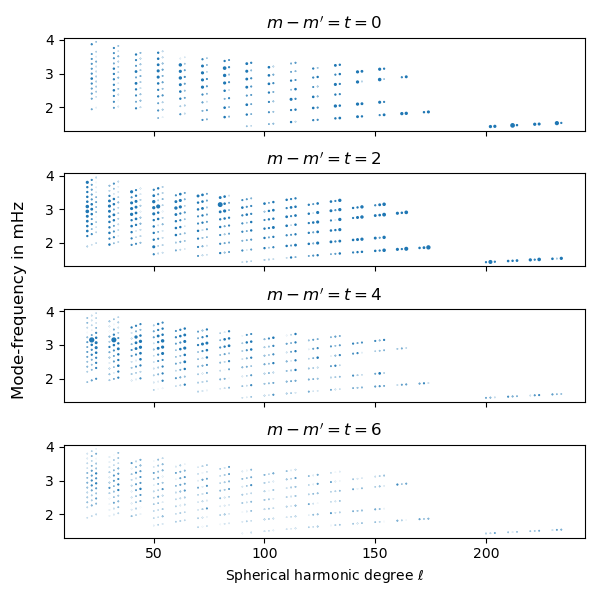}
  \caption{The change in reduced-$\chi^2$ computed for the imaginary part of cross-spectra with a chosen model for C2L. The spherical harmonic degree is indicated on the x-axis and the mode-frequency is indicated on the y-axis. The cross-spectra are computed for $\ell, \ell+\Delta\ell$ pairs for $\Delta\ell=0, 2, 4$ for $\ell=20, 30, 40...220, 230$. Groups of three vertical lines can be seen in the scatter plot. These lines correspond to ($\ell$---$\ell$), ($\ell$---$\ell+2$), ($\ell$---$\ell+4$) couplings respectively. The top panel has only two lines, as the power spectrum at $t=0$ is purely real. The size of the point indicates the magnitude of change.}
  \label{fig:misfit-t0246}
\end{figure}

\begin{figure}[!ht]
    \centering
    \includegraphics[width=0.9\textwidth]{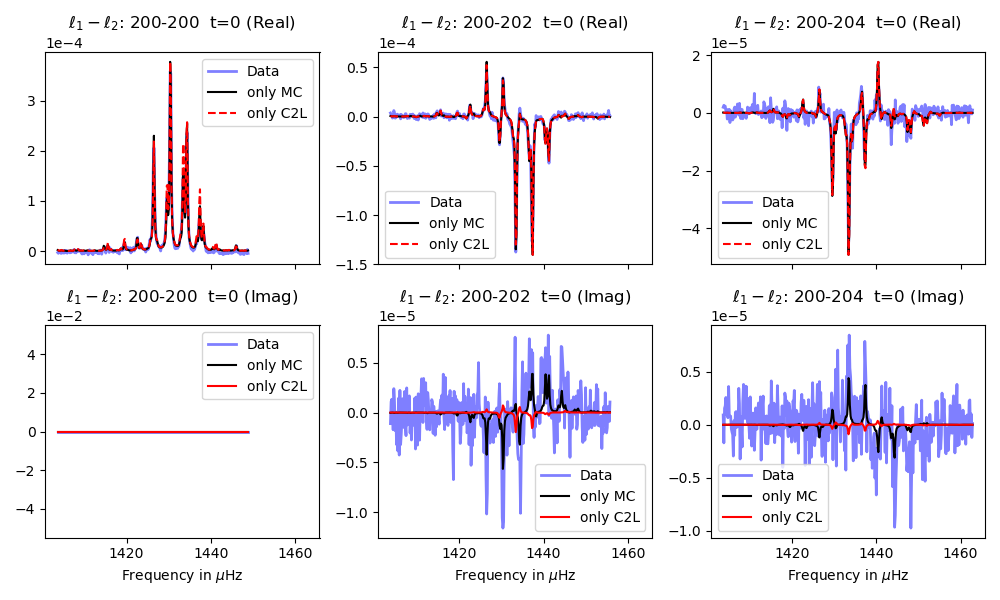}
    \caption{Cross-spectrum of modes for $n=0$. The spherical harmonic degrees of the coupled modes $\ell_1, \ell_2$ are indicated in the title of the sub-plot. The top-panels show the real part of the cross-spectra and the bottom panels show the imaginary parts of the cross-spectra. The blue lines indicate observed cross-spectra from HMI. The black lines indicate the cross-spectral model with differential rotation and meridional circulation profiles. The red lines mark model cross-spectra with rotation and a C2L model, without meridional circulation.}
    \label{fig:p4-cs-0-t0}
\end{figure}

We construct forward models for cross-spectra for different $\ell-\ell'$ pairs at the same radial order $n$. To understand and quantify the change in cross-spectra due to the C2L-model, we compute an L2-norm misfit function similar to that used by \cite{SGK-2021-ApJS}, with the addition that the misfit now includes both real and imaginary components of cross-spectra. We have
\begin{equation}
  D_{nt}^{\ell, \ell'}(\omega) = \left\langle \sum_m \mcS_m [\varphi^{\ell m}(\omega)\varphi^{\ell', m+t*}(\omega)]_\mathrm{data} \right\rangle
\end{equation}
\begin{equation}
  M = \sum_\omega \left[\mathrm{Re}[D^{\ell\ell'}_{nt}(\omega) - \mcM^{\ell\ell'}_{nt}(\omega)]\right]^2/\mathrm{Re}[\sigma^{\ell\ell'}_{nt}(\omega)] +
  \left[\mathrm{Im}[D^{\ell\ell'}_{nt}(\omega) - \mcM^{\ell\ell'}_{nt}(\omega)]\right]^2/\mathrm{Im}[\sigma^{\ell\ell'}_{nt}(\omega)],
  \label{eqn:p4-misfit}
\end{equation}
where $\sigma^{\ell\ell'}_{nt}(\omega)$ is the variance of the observed stack-summed cross-spectrum. Forward models are constructed for a wide range of modes across spherical harmonic degree $\ell$ and frequency, as shown in Figure~\ref{fig:nu-by-ell} -- the modes which propagate only in the convection zone are indicated with red, blue and black, while the modes which have finite amplitude in the radiative interior are shown in green. To demonstrate the importance of C2L modelling, we consider a model for the imaginary part of the spectrum for $\ell, \ell'=200, 202$ and $t=0$. We construct a 2-parameter C2L model with non-zero $g_1, g_2$ coefficients. We compute the misfit $M$ on a grid of values. Figure~\ref{fig:chisq-grid} shows that the optimal solution corresponds to a non-zero value of $g_1, g_2$, which indicates that a simultaneous fitting of both MC and C2L needs to be performed.
To understand how the C2L model affects cross-spectra across spherical harmonic degree and mode frequencies, we study the sensitivity of the misfit $M$ with changes to a model of C2L, given by $g_1=-1, g_2=-0.5$. Figure~\ref{fig:misfit-t0246} shows the change in misfit for different channels of the cross-spectra ($t=0, 2, 4, 6$) and for three possible couplings for a given $\ell$, which are $\ell, \ell+2, \ell+4$. For the case of $t=0$, only 2 possible couplings $\ell+2, \ell+4$ are significant as the $\ell$---$\ell$ cross-spectrum for $t=0$ is the power-spectrum, which is real by defintion. We note that $t=0$ and $t=2$ channels are the most affected by the C2L-bias model. For the $t=0$ channel, $\Delta\ell=2$ seems to be more significantly affected by C2L than the $\Delta\ell=4$ channel. For the $t=2$ cross-spectral channel, $\Delta\ell=0, 2$ are predominantly affected for $\ell<100$, but $\Delta\ell=2, 4$ channels change more substantially for $\ell>100$. The cross-spectra that are most affected have frequencies in the vicinity of 3 mHz. For the $t=4$ cross-spectral channel, $\Delta\ell=2, 4$ are most impacted by C2L-bias and the effect is strong only for $\ell<100$. The effect of C2L on the$t=6$ channel is found to be negligible. An example stack-summed cross-spectrum is shown in Figure~\ref{fig:p4-cs-0-t0}. For illustration, cross-spectra for various modes across spherical harmonic degree and frequency are shown in Figures~\ref{fig:p4-cs-0-t2}--\ref{fig:p4-cs-9-t2} of Appendix~\ref{apdx:cs-plots}.
\subsection{Preliminary inversions}
\begin{figure}
    \centering
    \includegraphics[width=0.5\textwidth]{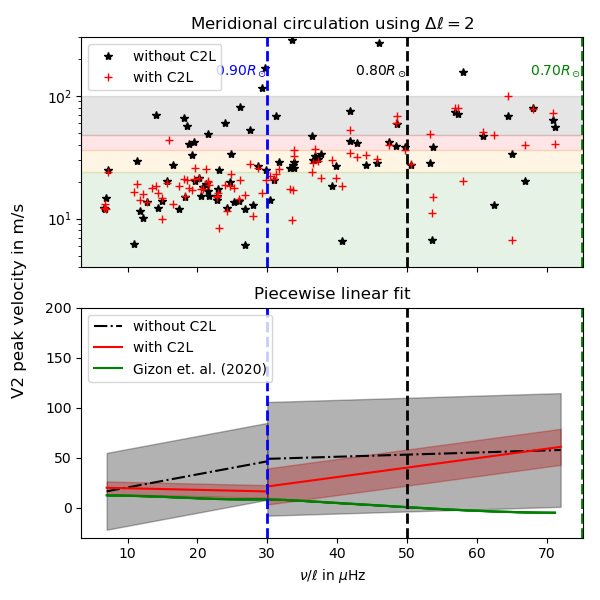}
    \caption{Peak vertical velocity $u_\mathrm{max}$ estimated from MC inferences using cross-spectra with $\Delta\ell=2$. Black stars correspond to a model without C2L and red crosses indicate that C2L has been accounted for. The bottom panel shows a linear fit to the points in the top panel in the two regions R1 and R2. The inversion result from \citetalias{Gizon-2020-Science} is shown in green in the bottom panel.}
    \label{fig:u2-dell2}
\end{figure}

\begin{figure}
    \centering
    \includegraphics[width=0.5\textwidth]{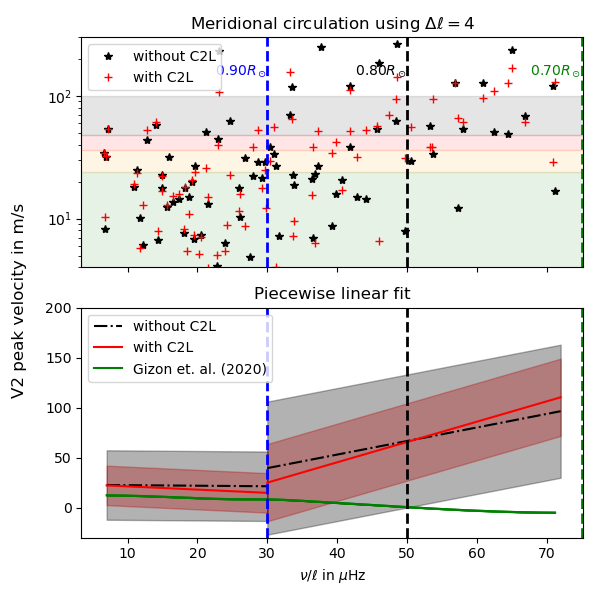}
    \caption{Peak vertical velocity $u_\mathrm{max}$ estimated from MC inference using cross-spectra with $\Delta\ell=4$. Black stars correspond to a model without C2L and the red crosses are when  C2L is taken into account. The bottom panel shows a linear fit to the points in the top panel in the two regions R1 and R2. The inversion result from \citetalias{Gizon-2020-Science} is shown in green in the bottom panel.}
    \label{fig:u2-dell4}
\end{figure}

To demonstrate the importance of including C2L-bias when modelling cross-spectra, we set up a preliminary inversion for the parameters $\mathrm{Im}[a_2^2]$, $\mathrm{Im}[a_2^4]$ (see Eqn.~\ref{eqn:ask-imag}) along with $\Lambda(r_H)$. We perform this inversion for each spectrum separately. Thus, for this demonstration, the frequency dependence of $\Lambda(r_H)$ is implicitly captured. We reproduce the calculations of \citetalias{Woodard-2013-SoPh} of MC estimation, with and without the C2L model. Figures \ref{fig:u2-dell2} and \ref{fig:u2-dell4} show the peak surface velocity $u_\mathrm{max}$ of an artificial flow $v_2(r)/r = \langle v_2(r)/r \rangle_{n\ell}$. Similar to the observations of \citetalias{Woodard-2013-SoPh}, these inversions may be broadly divided into 4 regions: modes with $\nu/\ell<30 \mu$Hz, whose lower turning points $r_\mathrm{LTP}>0.9 R_\odot$ (region R1), region R2 corresponding to $0.8 R_\odot \le r_\mathrm{LTP} \le 0.9 R_\odot$, region R3 corresponding to $0.7 R_\odot \le r_\mathrm{LTP} \le 0.8 R_\odot$ and another region which contains modes that propagate much deeper (region R4). In R1, we find that inclusion of C2L qualitatively changes the nature of inferred MC. However, the effect is not so strong for the deeper layers. We consider the results of \citetalias{Gizon-2020-Science} as a reference for comparison. As seen in Figure~\ref{fig:u2-dell2}, the inverted MC shows an opposite trend to the profile inferred by \citetalias{Gizon-2020-Science} if the spectra was fitted with an MC profile without modelling C2L. Introducing the C2L results in a close match in the depth variation of MC, albiet a slight over-estimation. For the deeper layers, the unphysical increase in estimated flows from cross-coupling measurements have also been previously observed (\citetalias{Woodard-2013-SoPh}). Inclusion of C2L not only reverses the trend in the shallow layers, it also reduces the scatter in the estimation. The change in depth variation is more drastic in the inference of MC from $\Delta\ell=2$ spectra, as compared to $\Delta\ell=4$. The computation was repeated using different initial guesses to establish robustness of the results of optimization.

\begin{figure}
    \centering
    \includegraphics[width=\textwidth]{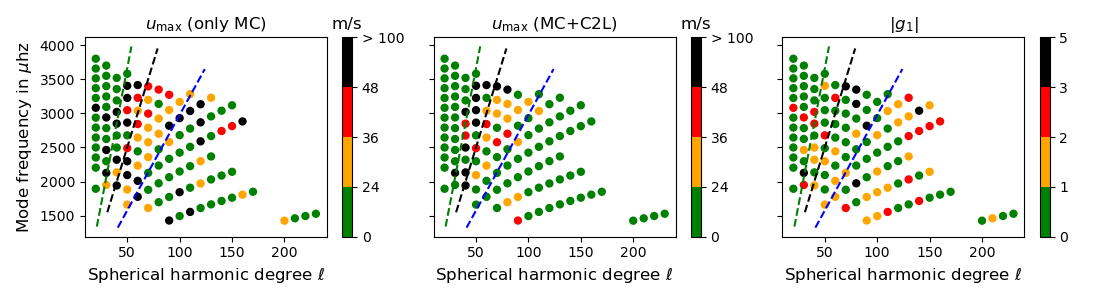}
    \caption{Peak vertical velocity $u_\mathrm{max}$ estimated using $\Delta\ell=2$ on the left panel and the $|g_1|$ coefficient of C2L systematic on the right panel. The middle panel shows $u_\mathrm{max}$ when both MC and C2L are fitted.}
    \label{fig:grid-plots-dell2}
\end{figure}
\begin{figure}
    \centering
    \includegraphics[width=\textwidth]{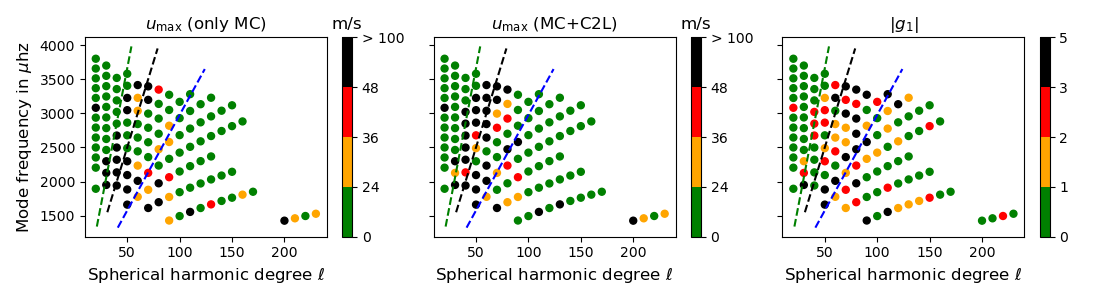}
    \caption{Peak vertical velocity $u_\mathrm{max}$ estimated using $\Delta\ell=4$ on the left panel and the $|g_1|$ coefficient of C2L systematic on the right panel. The middle panel shows $u_\mathrm{max}$ when both MC and C2L are fitted.}
    \label{fig:grid-plots-dell4}
\end{figure}
\subsection{Inversions for deeper layers}
\begin{figure}
    \centering
    \includegraphics[width=\textwidth]{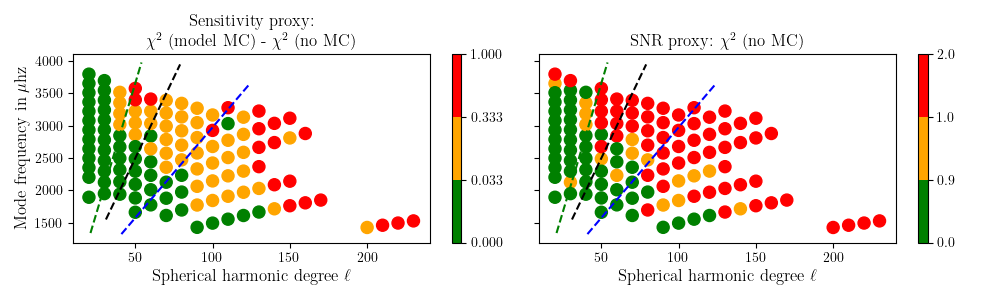}
    \caption{Proxy for model sensitivity to MC is shown on the left panel and proxy for SNR is shown on the right panel. The red dots have the highest SNR and are predominantly found when $\nu/\ell<30 \mu$Hz (region to the right of the blue line, see Figure~\ref{fig:nu-by-ell}). The green dots have the lowest SNR and apply modes with $\nu/\ell>30 \mu$Hz (region to the left of the blue line).}
    \label{fig:snr-proxy}
\end{figure}

Preliminary inversions seem to perform well for shallow depths when the C2L effect is included in the model, but show only marginal improvement for deeper layers. This may be attributed to the decreased sensitivity of the model to flow perturbations and a reduction in SNR in the observed spectra. The changes in the cross-spectrum are captured by the coupling coefficients $c_\ell^{\ell+p}$, which in turn depend on coefficients $b_k$, given by Eqns.(\ref{eqn:b_k_real}, \ref{eqn:b_k_imag}). Both these coefficients have a prefactor $\ell (\partial \tilde\omega/\partial \ell)^{-1}$, which decreases rapidly with $\ell$ \citep[see Figure 7 of][]{SGK-2021-ApJS}. The sensitivity may also be demonstrated by studying the misfit for different $\ell$, with and without an MC model. As seen on the left panel of Figure~\ref{fig:snr-proxy}, the low-$\ell$ spectra suffer from poor sensitivity to MC. To understand the reliability of inversions in the deeper layers, we plot the reduced-$\chi^2$ of the observed spectrum. If this quantity is greater than 1, then the individual spectra contain information of MC. As seen in Figure~\ref{fig:snr-proxy}, the SNR is generally high for modes with $\nu/\ell<30 \mu$Hz as well as for modes with $\ell>50$ and $\nu_{n\ell}>2700 \mu$Hz. Figures~\ref{fig:grid-plots-dell2}, \ref{fig:grid-plots-dell4} show the results of the inversion on a grid. Introduction of C2L results in decrease in magnitude of estimated MC in R1 and R2 when the $\Delta\ell=2$ spectra are fitted. Fitting spectra in R3 results in unphysically large estimates of MC, irrespective of whether C2L is included or not. The estimated MC values are very low in general in R4. The spurious high MC estimated in R4 for $\Delta\ell=2$ spectra behave better when C2L is introduced. Similar observations are harder to make for $\Delta\ell=4$ spectra as the SNR is poorer. A simultaneous inversion of a large number of spectra within a frequency window could potentially help mitigate this effect and provide us with reliable estimates of both MC and C2L.
%

\section{Inversions: Feasibility and Recipe}
We saw in the preceding section that MC, albeit a weak flow, leaves a measurable imprint on cross-spectral measurements for different $\Delta\ell, t$ values and modelling C2L does help improve the estimation of MC. However, the SNR is poor, especially for modes which propagate in the deeper layers of the Sun. One way to do this is to determine C2L - for some range of frequencies - using purely shallow modes and assume that it is fixed for all modes in that frequency range. Alternately, the inference could be improved by setting up a full inversion that simultaneously uses all the cross-spectral data. While the latter inversion is not performed here, we perform a simple analysis to understand the feasibility of such an inversion and recommend modelling simplifications that could help the inversion.\\

Ideally, the full inversion would involve an estimation of MC while simultaneously fitting for both real and imaginary parts of all cross-spectra (all possible allowed $\ell, \ell', m, m'$ combinations). Among a larger set of cross-radial-order and cross-harmonic-degree couplings, we consider the limited set for each mode, $(\ell, m)$, coupled to $(\ell, \ell+2, \ell+4), (m, m+1, m+2, m+3, m+4)$. Hence, for each mode $\ell, m$, we have $3*5=15$ different spectra. These spectra have real and imaginary-parts and have different positive and negative $m-$branches -- a total of $60$ spectra per mode. Accounting for a total of about 1500 observed modes $(n, \ell)$, we have $90,000$ different cross-spectra. An iterative inversion procedure would need to compute $90k$ spectra for every iteration -- estimated to take about 30 minutes on a 100-core compute node. Although the number of spectra is large, it is feasible to perform such an inversion on present-day compute nodes.\\

Given that performing a full inversion is feasible, we propose the following recipe for carrying out the inversion.

\begin{enumerate}
  \item Parameterize the C2L systematic using $\Lambda_s (\omega)$, since it is known that the C2L profile is a strong function of frequency \citep{Chen-Zhao-2018-ApJ, Rajaguru-Antia-2020-ASSP}. Divide the frequency range where most acoustic modes are found (2.5 - 4.5 mHz) into $k$ frequency bins, and the C2L estimate is considered to be constant over the bin. Based on the results of \cite{Rajaguru-Antia-2020-ASSP}, we recommend $k=20$ to be a reasonable number to capture the variation of C2L-bias
  as a function of frequency $\omega$. For each frequency bin, there are $s_\mathrm{max}$ unknowns. Based on the calculations above, $s_\mathrm{max} = 7$ is sufficient to capture the spatial variation of C2L-bias. This results in $k s_\mathrm{max} = 20 \times 7 = 140$ parameters for expressing the C2L-bias.
  \item Express the depth dependence of meridional circulation using B-splines
  (e.g., \cite{SBD-2023-ApJS} performed an inversion for rotation using B-splines that
  led to excellent reproduction of well-established rotation profiles.) This
  compact notation would enable the depth variation in the entirety of convection zone using only a handful of B-spline coefficients. Since MC is equatorially anti-symmetric, the selection rules ensure that
  only even-spherical harmonic degrees are needed to represent the flow. For representing the flow in the convection zone, we could use $50$ spline coefficients for each spherical harmonic degree $s=2, 4, 6$. Hence a total of $150$ parameters may be sufficient to represent the MC profile in the convection zone.
  \item Estimate MC and the C2L systematic by simultaneously fitting for
  both the real and imaginary parts of $90k$ spectra (all possible allowed $\ell, \ell', m, m'$ combinations). These estimates only consider couplings at the same radial order $n$. Considering cross-radial order couplings could result in a significant addition to the number of cross-spectra.
\end{enumerate}
Since the SNR for the imaginary part of the cross-spectrum is much lower than the real part,
using all the available data becomes a necessity, which makes setting up
the inversion an elaborate process, a challenge we will pursue in a future publication.
\begin{figure}
    \centering
    \includegraphics[width=0.7\textwidth]{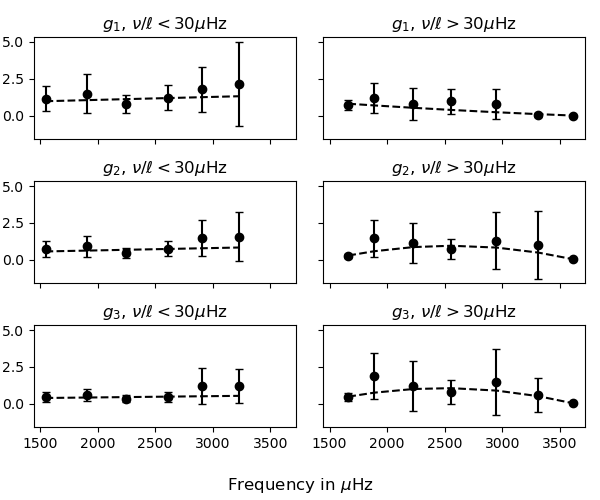}
    \caption{Variation of the C2l parameters as a function of frequency. The left panels correspond to values of C2L for modes with $\nu/\ell < 30 \mu$Hz and right panels correspond to the values of C2L parameters for modes with $\nu/\ell > 30\mu$Hz. The parameters for the shallow modes have a smaller spread.}
    \label{fig:c2l-freq-dependence}
\end{figure}
Figure~\ref{fig:c2l-freq-dependence} shows the results of binning the C2L parameters in different regions. We perform the binning in frequency for R1 and R2-R4 combined. After binning, we fit a quadratic polynomial as a function of frequency, in order to capture the frequency dependence of the C2L parameters. The parameters corresponding to C2L show a variation in frequency which depends the lower turning point of the modes in question in Figure~\ref{fig:c2l-freq-dependence}. For deeper-propagating modes, $g_j$ appears to show a trend which grows with frequency, whereas the trend is different for the shallow modes. These trends are not statistically significant and are therefore inconclusive. The trends could only be reliably established by performing a more elaborate fitting as described earlier.

\section{Discussions}
\label{sec:discussions}
The depth profile of MC has been a topic of debate for decades. The
majority of MC inferences has been from techniques such as time-distance helioseismology and ring-diagram analysis, necessitating the exploration of other methods and approaches to build consensus.
Owing to its utilization of complete modal
information, normal-mode coupling \citep{Lavely-Ritzwoller-1992-RSPTA, Hanasoge-2017-MNRAS, SGK-2021-ApJS}
promises to be a technique which has the potential to provide high-quality inferences
of the depth profiles of flows. Present-day computers enable the use of
mode-coupling techniques since the computation of a large
number of cross-spectra may be rapidly accomplished. \\

While the role of the C2L systematic in complicating MC estimation is
well documented in the context of time-distance helioseismology \citep{Zhao-2012-ApJL, Zhao-2016-SoPh}, efforts to understand its effect in the context of mode-coupling have been limited. While \citetalias{Woodard-2013-SoPh} highlight the possibility of C2L bias to be crucial for estimation of meridional circulation, \cite{Schad-2013-ApJL} predict a multi-cell meridional circulation and claim the effect of C2L-bias to be negligible for their cross-coupling measurable for $\ell < 100$. 
The preliminary inversions presented here also demonstrate that C2L may have a strong influence on lower $\ell$ modes. Such a model seems to improve MC estimation in shallow layers, but does not help us rule out the presence of other systematics which could affect the deeper layers. Particularly, the competition between C2L and MC for defining the line profile of cross spectra at low $\ell$ could be resolved by performing inversions from the entire cross-spectral dataset. Normal mode coupling has recently been used to detect and measure Rossby and inertial modes \citep{Mandal-2020-ApJ, Mandal-2024-arXiv} and holds promise for being able to characterize all internal time-varying flows. Hence it is important to model the C2L-bias in the context of normal mode coupling, especially while attempting to measure weak flows such as meridional flow. However, computing these cross-spectra are computationally more expensive. Since the C2L-bias is modelled in the form of a phase factor and MC affects the imaginary part of the cross-spectrum,  calculations need to be performed with complex datatypes. This results in the slow-down of the computation of each cross-spectrum by a factor of 16. We propose the usage of the entire set of spectra for constraining MC. 72-day time series' from HMI and MDI typically resolve between 1500-1700 modes (unique $(n, \ell)$). As shown in the calculation in the previous section, the information that could be used to constrain MC would be $90,000$ cross-spectra. This renders MCMC methods, such as the ones used by \cite{SGK-2021-ApJS} for the estimation of rotation, unrealistic for the MC problem. However, traditional inversion methods could still be employed for the estimation of parameters. Fast-gradient computation tools such as autograd \cite{Maclaurin-2015-github} enable an efficient computation of numerical gradients, thus making their use attractive for inversion methods, which usually involve the computation of gradients and Hessians. The total computation time for each spectrum on a single core is $\sim$ 2 seconds. Hence, inversion methods are tractable on 100-core compute clusters. Given that we now have 13 years of continuous observations from HMI, averaging a large number of 72-day time-series segments could be used to significantly improve SNR and potentially provide an alternate and robust inference of time-averaged MC.
\section*{Acknowledgements}
We thank Jesper Schou, Max Planck Institute for Solar System Research, G\"{o}ttingen, for discussions and for sharing the proceedings of \cite{Schou-2013-enss}. We thank the anonymous reviewer for their helpful comments. The work was carried out as part of SGK's Ph. D. thesis at TIFR. \\
\newline
\textit{Software}: The codes used the following Python libraries: Numpy \citep{numpy} and matplotlib \citep{matplotlib} 

\clearpage
\begin{appendix}

\section{Vector Spherical Harmonics}
\label{apdx:vsh}
Vector spherical harmonic components are given by
\begin{align}
 \bfY^{\ell m}(\theta, \phi) & \equiv \hat\bfr \, Y^{\ell m}(\theta, \phi) \\
 \bfPsi^{\ell m}(\theta, \phi) & \equiv \bfnabla_h Y^{\ell m}(\theta, \phi) \\
 \bfPhi_{\ell m}(\theta, \phi) & \equiv \hat\bfr \times
                                \bfnabla_h Y^{\ell m}(\theta, \phi),
\end{align}
where $Y^{\ell m}(\theta, \phi)$ are spherical harmonics and $\hat\bfr$ is the
radial unit vector and $\bfnabla_h$ is the horizontal gradient operator
given by
\begin{equation}
 \bfnabla_h = \hat{\bftheta}\frac{\partial}{\partial \theta}
                + \hat{\bfphi}\frac{1}{\sin\theta}\frac{\partial }
                {\partial \phi}.
\end{equation}
Vector spherical harmonics are orthogonal. The orthogonality
may be expressed in compact form if we define \(\bfGamma_0^{\ell m}
\equiv \bfY^{\ell m}\), \(\bfGamma_1^{\ell m} = \bfPsi^{\ell m}\) and
\(\bfGamma_2^{\ell m} = \bfPhi^{\ell m}\).
\begin{align}
  \int_\odot \rmd\Omega \; \bfGamma_i^{\ell m} \cdot \bfGamma_j^{*\ell' m'}
  = N_i^\ell \delta_{ij} \delta^{\ell \ell'} \delta^{mm'},
\end{align}
where \(N_{il}\) is a normalization constant,
\(N_0^\ell = 1, N_1^\ell = N_2^\ell = \ell(\ell+1)\) and
\(\rmd\Omega = \sin\theta \, \rmd\theta \, \rmd\phi\) is the surface element,
integration being performed over the entire surface of the sphere, denoted by $\odot$.
\section{Spherical harmonics symmetry relations}
\label{apdx:sph-symm}
Consider a time-varying, real-valued scalar field on a sphere
\(\varphi(\theta, \phi, \tau)\). The
spherical-harmonic components are given by
\begin{equation}
    \varphi^{\ell |m|}(t) = \int_\Omega \rmd\Omega \; Y^{*\ell |m|}(\theta, \phi) \varphi(\theta, \phi, \tau)
    = (-1)^{|m|} \int_\Omega \rmd\Omega \; Y^{\ell, -|m|}(\theta, \phi) \varphi(\theta, \phi, \tau) = (-1)^{|m|} \varphi^{*\ell, -|m|}(t),
\end{equation}
where $\rmd\Omega$ is the area element, the integration being performed over
the entire surface of the sphere.
After performing a temporal Fourier transform, we have
\begin{equation}
    \varphi^{\ell |m|}(\omega) = \frac{1}{\sqrt{2\pi}}\int_{-\infty}^\infty \rmd t
    e^{-i\omega t} \varphi^{\ell |m|}(t)
    = (-1)^{|m|} \frac{1}{\sqrt{2\pi}}\int_{-\infty}^\infty \rmd t
    e^{-i\omega t} \varphi^{* \ell, -|m|}(t),
\end{equation}
\begin{equation}
\varphi^{*\ell, -|m|}(\omega) = \frac{1}{\sqrt{2\pi}} \int_{-\infty}^\infty \rmd t e^{i\omega t}
\varphi^{*\ell, -|m|}(t) = (-1)^{|m|} \varphi^{\ell |m|}(-\omega) \implies
\varphi^{\ell, -|m|}(\omega) = (-1)^{|m|} \varphi^{*\ell |m|}(-\omega).
\end{equation}
\section{Derivation of the modified leakage matrix}
\label{sec:deriv-leakage}
Using Eqns.~(\ref{eq:true-wavefield-sht}, \ref{eq:wavefield-sht}),
we express the spherical-harmonic components of the observed
wavefield $\varphi^{\ell m}$ in terms of their ``true" counterparts
$\varPhi^{\ell' m'}$ and the C2L-bias components $\Lambda^{st}$ thus
\begin{align}
    \varphi^{\ell m}(t) & = \sum_{\ell' m' s t} \varPhi^{\ell' m'}(t)
    \Lambda^{st}
    \int_\odot d\Omega \, Y^{\ell' m'} Y^{st} Y^{\ell m*} \nonumber \\
    & = (-1)^{m} \sum_{\ell' m' s t} \varPhi^{\ell' m'}(t) \Lambda^{st}
    \int_\odot d\Omega \, Y^{\ell' m'} Y^{st} Y^{\ell, -m}.
\end{align}

The integral of a triple product of spherical harmonics over the entire surface may be written in terms of a product of Wigner-3j symbols,
\begin{align}
    \varphi^{\ell m}(t) & = (-1)^{m} \sum_{\ell' m' s t}
     \Lambda^{st} \gamma_\ell \gamma_s \gamma_{\ell'}
     \wig{\ell'}{s}{\ell}{0}{0}{0}
     \wig{\ell'}{s}{\ell}{m'}{t}{-m} \varPhi^{\ell' m'}(t) \nonumber \\
    \varphi^{\ell m}(t) & = (-1)^{m} \gamma_\ell \sum_{\ell' s}
       \gamma_s \gamma_{\ell'}
     \wig{\ell'}{s}{\ell}{0}{0}{0}
     \sum_{m' t}\Lambda^{st}
     \wig{\ell'}{s}{\ell}{m'}{t}{-m} \varPhi^{\ell' m'}(t).
\end{align}
Since all factors other than the wavefield are independent of time, the
same relation holds even in frequency domain,
\begin{equation}
    \varphi^{\ell m}(\omega) = (-1)^{m} \gamma_\ell \sum_{\ell' s}
       \gamma_s \gamma_{\ell'}
     \wig{\ell'}{s}{\ell}{0}{0}{0}
     \sum_{m' t}\Lambda^{st}
     \wig{\ell'}{s}{\ell}{m'}{t}{-m} \varPhi^{\ell' m'}(\omega).
\end{equation}

Modelling the wavefield using the existing leakage matrices, we have
\begin{equation}
    \varPhi^{\ell m}(\omega) = \tilde{L}^{\ell m}_{\ell_1 m_1}
    c^{\ell_1 m_1}_{\ell_2 m_2}
    a^{\ell_2 m_2}(\omega),
\end{equation}

\begin{equation}
    \varphi^{\ell m}(\omega) = (-1)^{m} \gamma_\ell \sum_{\ell' s}
       \gamma_s \gamma_{\ell'}
     \wig{\ell'}{s}{\ell}{0}{0}{0}
     \sum_{m' t}\Lambda^{st}
     \wig{\ell'}{s}{\ell}{m'}{t}{-m}
     \tilde{L}^{\ell m}_{\ell_1 m_1} c^{\ell_1 m_1}_{\ell_2 m_2}
     a^{\ell_2 m_2}(\omega)
     = L^{\ell' m'}_{\ell_1 m_1} c^{\ell_1 m_1}_{\ell_2 m_2}
     a^{\ell_2 m_2}(\omega).
\end{equation}
Hence, the modified leakage matrices $L$ may be written in terms of
the existing leakage matrices $\tilde{L}$ as
\begin{equation}
    L^{\ell m}_{\ell_1 m_1} = (-1)^{m} \gamma_\ell \sum_{\ell' s}
       \gamma_s \gamma_{\ell'}
     \wig{\ell'}{s}{\ell}{0}{0}{0}
     \sum_{m' t}\Lambda^{st}
     \wig{\ell'}{s}{\ell}{m'}{t}{-m}
     \tilde{L}^{\ell' m'}_{\ell_1 m_1}
     \label{eq:deriv-leakage}.
\end{equation}

\section{Parameterization of C2L bias}
\label{sec:apdx:parameterization}
The C2L systematic may be modelled as a feature which is axisymmetric in
a reference frame where the axis passes through the center of the
disk. We decompose the C2L profile in spherical harmonics, noting that $\Lambda_{st} = 0$ for
$t\neq 0$. For the present purpose, we assume that the C2L profile does not have rapid spatial variations and may thus be expressed using only a few low-degree spherical harmonics, i.e., $\Lambda_{st}=0$ for $s>s_\mathrm{max}$. The C2L-profile components that modify the leakage matrix need to be expressed in the standard heliographic coordinate frame. Hence, we need to perform a rotation of the spherical-harmonic components before using them to modify the leakage matrices, given by \citep[C255;][]{Dahlen-Tromp-1998-Book}
\begin{equation}
    d^{(\ell)}_{m' m}(\beta) = P^{m'}_{\ell m}(\cos\beta),
\end{equation}
where $P^N_{\ell m}$ correspond to generalized Legendre polynomials,
\citep[C112;][]{Dahlen-Tromp-1998-Book}. Spherical harmonics in
the rotated frame are described by
\begin{equation}
\bfY^{'N}_{\ell m} = \mcD^{\ell}_{m' m}(\alpha, \beta, \gamma)
\bfY^{N}_{\ell m'} =
\exp(i m' \gamma) d^{(\ell)}_{m' m}(\beta) \exp(im\alpha)
\bfY^{N}_{\ell m'},
\label{eq:sph-rotation}
\end{equation}
where $(\alpha, \beta, \gamma)$ correspond to the Euler angles of
rotation. In our case, we are dealing with scalar spherical harmonics and hence
$N=0$. The rotation may be  described by the Euler
angle $(0, \pi/2, 0)$.
Denoting $\bfY$ at the disk-center by $\bfY^\mathrm{(DC)}$ and the
$\bfY$ at the solar-north by $\bfY^\mathrm{(S)}$ greatly simplifies the rotation transformation,
\begin{equation}
\bfY^\mathrm{(S)}_{\ell m} = d^{(\ell)}_{m' m}(\beta) \bfY_{\ell m'}
= P^{m'}_{\ell m}(\cos\beta) \bfY^\mathrm{(DC)}_{\ell m'}.
\label{eq:ylm-rotation}
\end{equation}
Similarly, the coordinate transformation may also be performed in the opposite
sense, i.e., moving from the S frame to the DC frame,
\begin{equation}
\bfY^\mathrm{(DC)}_{\ell m} = d^{(\ell)}_{m' m}(-\beta)
\bfY^\mathrm{(S)}_{\ell m'}
= P^{m'}_{\ell m}(\cos\beta) \bfY^\mathrm{(S)}_{\ell m'}
\label{eq:ylm-rotation2}.
\end{equation}

The C2L profile, given by $\Lambda(r_\rmH)$, where $r_\rmH$ is the heliocentric radius,
is expanded in terms of spherical harmonics as
\begin{align}
\Lambda(r_\rmH) &= \sum_{st} \Lambda^\mathrm{(DC)}_{st}\bfY^\mathrm{(DC)}_{st}
= \sum_{s t'} \Lambda^\mathrm{(S)}_{s t'} \bfY^\mathrm{(S)}_{s t'} \\
\Lambda(r_\rmH) &= \sum_{s t'} \Lambda^\mathrm{(DC)}_{s t'} \sum_{t''} P^{t''}_{s t'}(\cos\beta) \bfY^\mathrm{(S)}_{st''} \\
\Lambda(r_\rmH) &= \sum_{st''}
\left( \sum_{t'} \Lambda^\mathrm{(DC)}_{st'} P^{t''}_{st'}(\cos\beta)
\right) \bfY^\mathrm{(S)}_{st''}.
\end{align}

This expression is used to transform the $\Lambda_s$ parameters to
the corresponding spherical-harmonic coefficients in the correct coordinate
frame, i.e.,

\begin{align}
\Lambda^\mathrm{(S)}_{st} = \sum_{t'} \Lambda^\mathrm{(DC)}_{st'} P^{t}_{st'}(\cos\beta) \\
\Lambda^\mathrm{(S)}_{st} = \Lambda^\mathrm{(DC)}_{s0} P^{t}_{s0}(\cos\beta)
\label{eq:sph-coeff-rot}.
\end{align}

Using the symmetry relations \citep[C.118;][]{Dahlen-Tromp-1998-Book}
$P^N_{\ell m}(\cos\beta) = (-1)^{N+m} P^m_{\ell N}(\cos\beta)$,
we obtain,

\begin{equation}
\Lambda^\mathrm{(S)}_{st} = (-1)^t \Lambda^\mathrm{(DC)}_{s0} P^0_{st}(\cos\beta).
\end{equation}

The generalized Legendre function $P^0_{st}$ can be written in terms of
associated Legendre polynomials, $P_{st}$ as \citep[C.113;][]{Dahlen-Tromp-1998-Book}
\begin{equation}
  P_{\ell m}^0 = (-1)^m \left[ \frac{(\ell - m)!}{(\ell + m)!}\right]^{1/2} P_{\ell m}.
\end{equation}

Hence, the transformation equation for $\Lambda_{st}$ can be written in terms of
associated Legendre polynomials $P_{st}$ as
\begin{equation}
\Lambda^\mathrm{(S)}_{st} = \left[\frac{(s - t)!}{(s + t)!} \right]^{1/2} \Lambda^\mathrm{(DC)}_{s0} P_{st}(\cos\beta).
\end{equation}

\subsection{Constraints on $\Lambda$}
Although $\Lambda_s$ is a free parameter of the problem, it represents the
spherical-harmonic components of a phase factor and is hence
constrained to be such that $|\Lambda(r_\rmH)|^2 = 1$ for all $r_\rmH$. This means that,
for all possible closed-surfaces on the sphere, which encloses a solid angle $\mcS$,
\begin{align}
|\Lambda(r_\rmH)|^2 & =
\left( \sum_s \Lambda_s \bfY_{s0} \right)
\left( \sum_s \Lambda^*_{s'} \bfY^*_{s' 0} \right) \\
& = \sum_{ss'} \Lambda_s \Lambda^*_{s'} \bfY_{s0} \bfY^*_{s'0}
\\
\implies \int_\mcS d\Omega \, |\Lambda(r_\rmH)|^2 &= \mcS =
\sum_{ss'} \Lambda_s \Lambda^*_{s'}
\int_\mcS d\Omega \, \bfY_{s0} \bfY^*_{s'0} = \sum_{s} |\Lambda_s|^2
\label{eq:constraints}
\end{align}
A weak form of the constraint can be obtained by considering the entire sphere, i.e.
$\mcS = 4\pi$. Defining $\Lambda'_s = \Lambda_s/\sqrt{4\pi}$, the constraint becomes
\begin{equation}
    \sum_{s} |\Lambda'_s|^2 = 1.
\end{equation}
Another condition may be derived by considering that the
phase factor is real. This is important as the imaginary component
of such a phase factor would contribute to growth/attenuation of
the spectrum in question. Hence,
\begin{equation}
 \Lambda(r_\rmH) = \exp(i f(r_\rmH)),
\end{equation}
where $f(r_\rmH)$ is real.
\section{Comparison of various cross-spectral models}
\label{apdx:cs-plots}
\begin{figure}[!ht]
    \centering
    \includegraphics[width=0.9\textwidth]{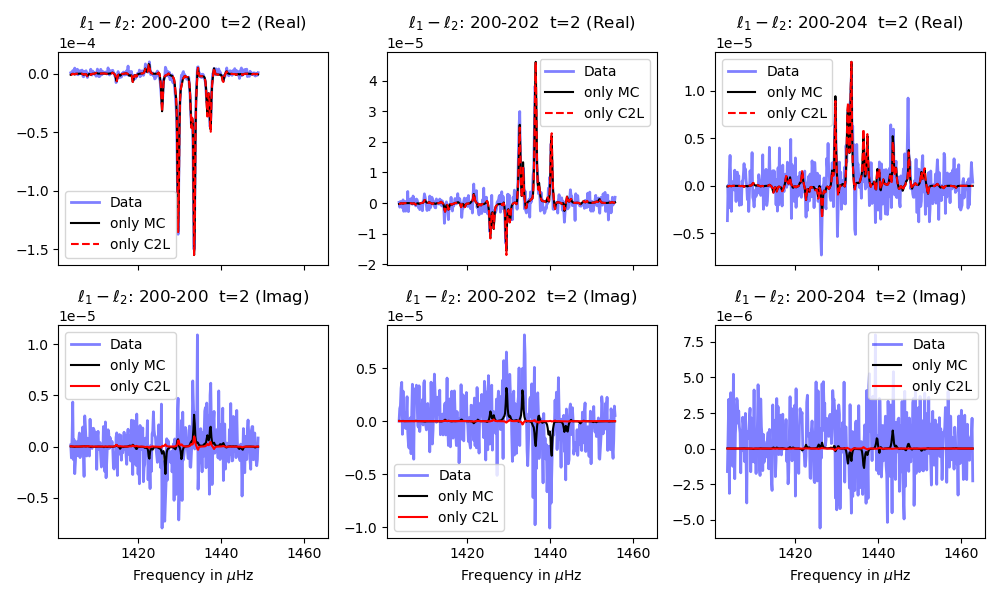}
    \caption{Cross-spectrum of modes for $n=0$ and $t=m_1-m_2=2$. The spherical harmonic degrees of the coupled modes $\ell_1, \ell_2$ are indicated in the title of the sub-plot. The top-panels show the real part of the cross-spectra and the bottom panels show the imaginary parts of the cross-spectra. The blue lines indicate observed cross-spectra from HMI. The black lines indicate cross-spectral model with differential rotation and meridional circulation profiles. The red lines indicate model cross-spectra with rotation and a C2L model, without meridional circulation.}
    \label{fig:p4-cs-0-t2}
\end{figure}
\begin{figure}[!ht]
    \centering
    \includegraphics[width=0.9\textwidth]{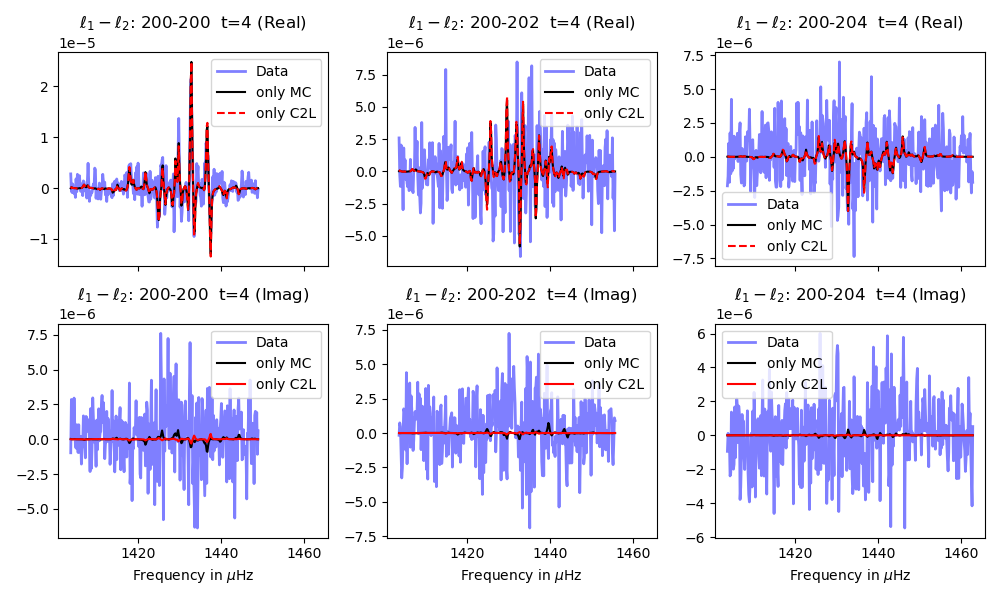}
    \caption{Cross-spectrum of modes for $n=0$ and $t=m_1-m_2=4$. The spherical harmonic degrees of the coupled modes $\ell_1, \ell_2$ are indicated in the title of the sub-plot. The top-panels show the real part of the cross-spectra and the bottom panels show the imaginary parts of the cross-spectra. The blue lines indicate observed cross-spectra from HMI. The black lines indicate cross-spectral model with differential rotation and meridional circulation profiles. The red lines indicate model cross-spectra with rotation and a C2L model, without meridional circulation.}
    \label{fig:p4-cs-0-t4}
\end{figure}
\begin{figure}[!ht]
    \centering
    \includegraphics[width=0.9\textwidth]{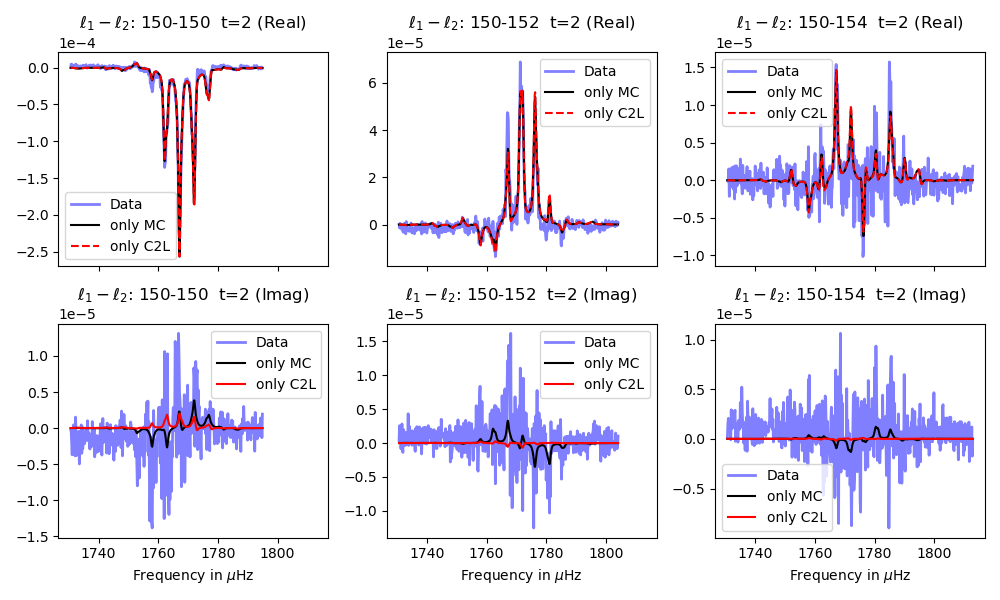}
    \caption{Cross-spectrum of modes for $n=1$ and $t=m_1-m_2=2$. The spherical harmonic degrees of the coupled modes $\ell_1, \ell_2$ are indicated in the title of the sub-plot. The top-panels show the real part of the cross-spectra and the bottom panels show the imaginary parts of the cross-spectra. The blue lines indicate observed cross-spectra from HMI. The black lines indicate cross-spectral model with differential rotation and meridional circulation profiles. The red lines indicate model cross-spectra with rotation and a 2L model, without meridional circulation.}
    \label{fig:p4-cs-1-t2}
\end{figure}
\begin{figure}[!ht]
    \centering
    \includegraphics[width=0.9\textwidth]{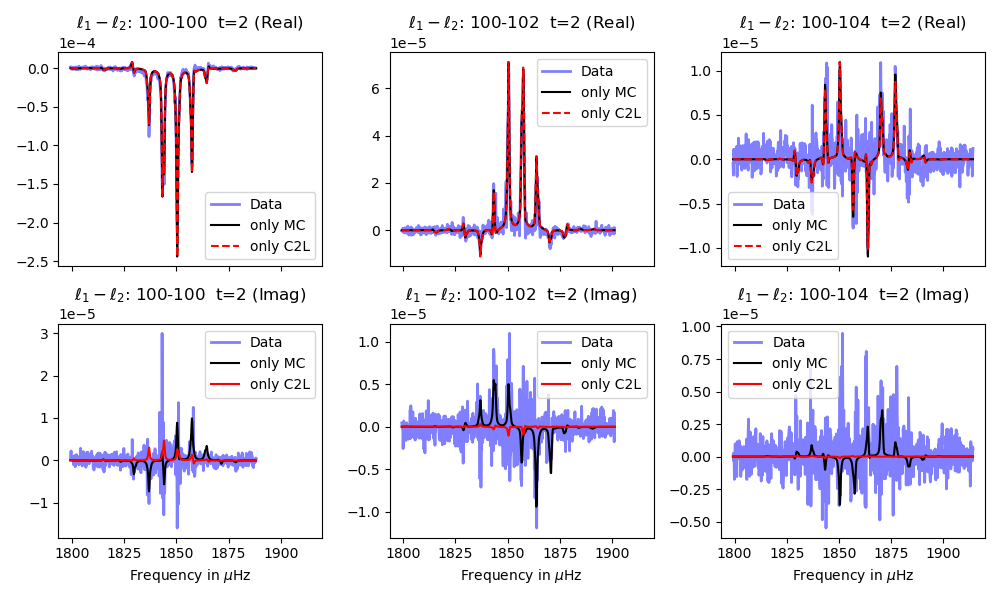}
    \caption{Cross-spectrum of modes for $n=2$ and $t=m_1-m_2=2$. The spherical harmonic degrees of the coupled modes $\ell_1, \ell_2$ are indicated in the title of the sub-plot. The top-panels show the real part of the cross-spectra and the bottom panels show the imaginary parts of the cross-spectra. The blue lines indicate observed cross-spectra from HMI. The black lines indicate cross-spectral model with differential rotation and meridional circulation profiles. The red lines indicate model cross-spectra with rotation and a C2L model, without meridional circulation.}
    \label{fig:p4-cs-2-t2}
\end{figure}
\begin{figure}[!ht]
    \centering
    \includegraphics[width=0.9\textwidth]{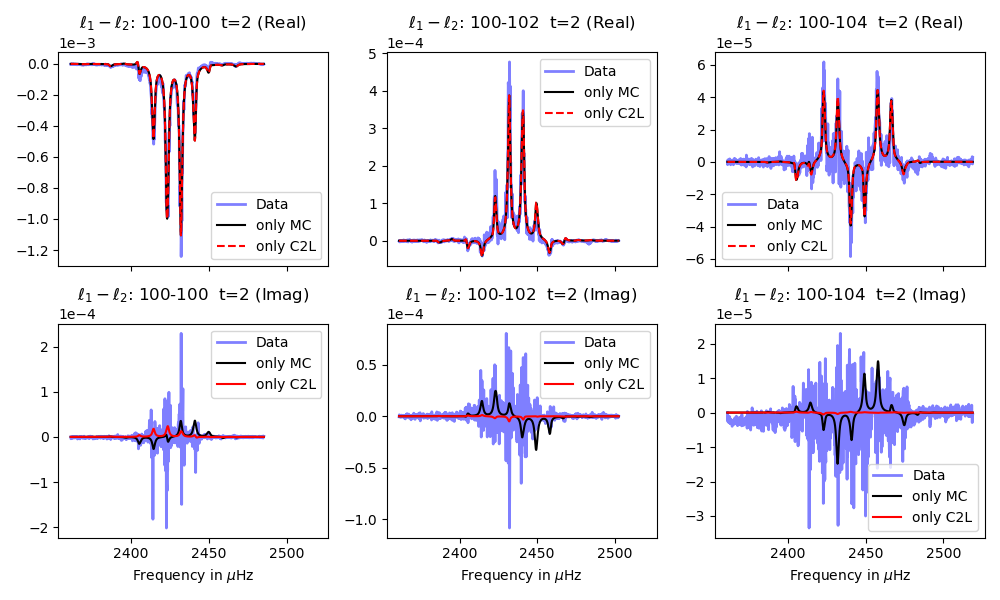}
    \caption{Cross-spectrum of modes for $n=4$ and $t=m_1-m_2=2$. The spherical harmonic degrees of the coupled modes $\ell_1, \ell_2$ are indicated in the title of the sub-plot. The top-panels show the real part of the cross-spectra and the bottom panels show the imaginary parts of the cross-spectra. The blue lines indicate observed cross-spectra from HMI. The black lines indicate cross-spectral model with differential rotation and meridional circulation profiles. The red lines indicate model cross-spectra with rotation and a C2L model, without meridional circulation.}
    \label{fig:p4-cs-4-t2}
\end{figure}
\begin{figure}[!ht]
    \centering
    \includegraphics[width=0.9\textwidth]{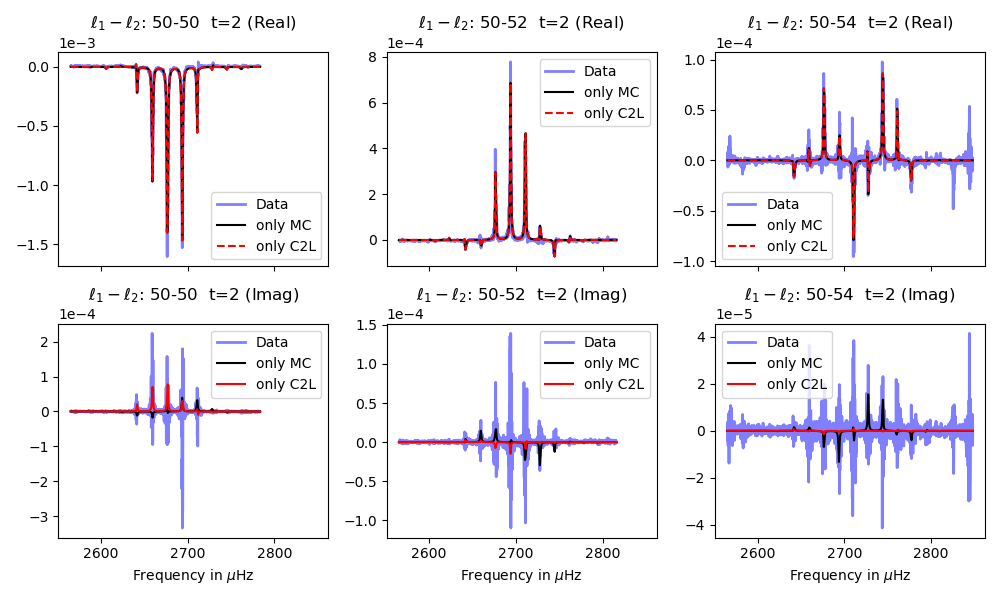}
    \caption{Cross-spectrum of modes for $n=8$ and $t=m_1-m_2=2$. The spherical harmonic degrees of the coupled modes $\ell_1, \ell_2$ are indicated in the title of the sub-plot. The top-panels show the real part of the cross-spectra and the bottom panels show the imaginary parts of the cross-spectra. The blue lines indicate observed cross-spectra from HMI. The black lines indicate cross-spectral model with differential rotation and meridional circulation profiles. The red lines indicate model cross-spectra with rotation and a C2L model, without meridional circulation.}
    \label{fig:p4-cs-8-t2}
\end{figure}
\begin{figure}[!ht]
    \centering
    \includegraphics[width=0.9\textwidth]{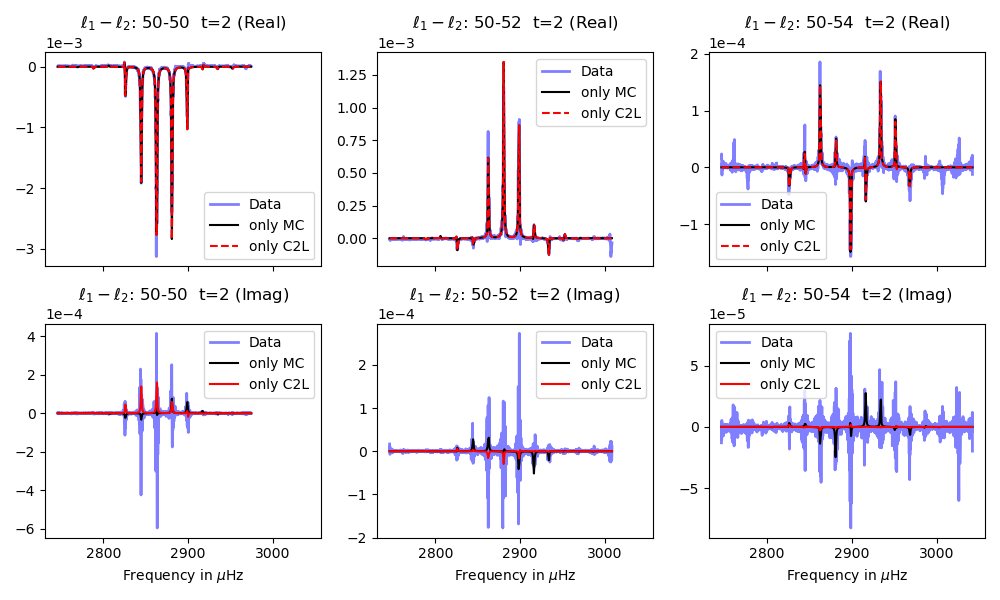}
    \caption{Cross-spectrum of modes for $n=9$ and $t=m_1-m_2=2$. The spherical harmonic degrees of the coupled modes $\ell_1, \ell_2$ are indicated in the title of the sub-plot. The top-panels show the real part of the cross-spectra and the bottom panels show the imaginary parts of the cross-spectra. The blue lines indicate observed cross-spectra from HMI. The black lines indicate cross-spectral model with differential rotation and meridional circulation profiles. The red lines indicate model cross-spectra with rotation and a C2L model, without meridional circulation.}
    \label{fig:p4-cs-9-t2}
\end{figure}

\end{appendix}


\newpage
\bibliography{references}{}
\bibliographystyle{aasjournal}


\end{document}

%% file: newcommandstex
\newcommand{\bftheta}{{\boldsymbol{\theta}}}

\newcommand{\bfxi}{{\boldsymbol{\xi}}}

\newcommand{\bfphi}{{\boldsymbol{\phi}}}

\newcommand{\bfnabla}{{\boldsymbol{\nabla}}}
\newcommand{\rmH}{\mathrm{H}}

\newcommand{\bfGamma}{{\boldsymbol{\Gamma}}}

\newcommand{\bfPhi}{{\boldsymbol{\Phi}}}
\newcommand{\bfPsi}{{\boldsymbol{\Psi}}}

\def\rmd{{\mathrm{d}}}

\newcommand{\tdot}{\,.\hspace{-0.98 mm}\raise.6ex\hbox{.}
                   \hspace{-0.98 mm}\raise1.2ex\hbox{.}\,}

\def\bfr{{\mathbf{r}}}

\def\bfu{{\mathbf{u}}}

\def\bfY{{\mathbf{Y}}}

\def\mcD{{\mathcal{D}}}

\def\mcM{{\mathcal{M}}}

\def\mcS{{\mathcal{S}}}

\newcommand{\wig}[6]{\begin{pmatrix}#1 & #2 & #3 \\#4 & #5 & #6\end{pmatrix}}